\documentclass[11pt,preprint]{aastex}
\def\lsim{\raise0.3ex\hbox{$<$}\kern-0.75em{\lower0.65ex\hbox{$\sim$}}}
\def\gsim{\raise0.3ex\hbox{$>$}\kern-0.75em{\lower0.65ex\hbox{$\sim$}}}

\begin{document}

\title{The Molecular Outflows in the $\rho$ Ophiuchi Main Cloud: \\
Implications For Turbulence Generation}
 
\author{Fumitaka Nakamura\altaffilmark{1,2},  
Yuhei Kamada\altaffilmark{3},
Takeshi Kamazaki\altaffilmark{1},  
Ryohei Kawabe\altaffilmark{4},
Yoshimi Kitamura\altaffilmark{2}, 
Yoshito Shimajiri\altaffilmark{4}, 
Takashi Tsukagoshi\altaffilmark{5},  
Kengo Tachihara\altaffilmark{1}, 
Toshiya Akashi\altaffilmark{6}, 
Kenta Azegami\altaffilmark{6}, 
Norio Ikeda\altaffilmark{2}, 
Yasutaka Kurono\altaffilmark{1}, 
Zhi-Yun Li\altaffilmark{7}, 
Tomoya Miura\altaffilmark{3}, 
Ryoichi Nishi\altaffilmark{3},
and Tomofumi Umemoto\altaffilmark{1}} 
\altaffiltext{1}{National Astronomical Observatory, Mitaka, Tokyo 181-8588, 
Japan; fumitaka.nakamura@nao.ac.jp}
\altaffiltext{2}{Institute of Space and Astronautical Science, 
Japan Aerospace Exploration Agency, 3-1-1 Yoshinodai, Sagamihara, 
Kanagawa 229-8510, Japan}
\altaffiltext{3}{Department of Physics, 
Niigata University, 8050 Ikarashi-2, Niigata, 950-2181, Japan}
\altaffiltext{4}{Nobeyama Radio Observatory, Nobeyama, 
Minamimaki, Minamisaku, Nagano, 384-1305, Japan}
\altaffiltext{5}{Department of Astronomy, 
School of Science, University of Tokyo,Bunkyo,
Tokyo 113-0033, Japan}
\altaffiltext{6}{
Department of Earth and Planetary Sciences, Tokyo Institute of 
Technology, Meguro, Tokyo 152-8551}
\altaffiltext{7}{Department of Astronomy, University of Virginia,
P. O. Box 400325, Charlottesville, VA 22904}

\begin{abstract}
We present the results of CO ($J=3-2$) and CO ($J=1-0$) mapping observations
toward the active cluster forming clump, L1688, in the $\rho$ Ophiuchi
 molecular cloud.
From the CO ($J=3-2$) and CO ($J=1-0$) data cubes, we identify 
five outflows, whose driving sources are VLA 1623, EL 32, LFAM 26, EL
 29, and IRS 44.
Among the identified outflows, the most luminous outflow 
is the one from the prototypical  Class 0 source, VLA 1623.  
We also discover that the EL 32 outflow located 
in the Oph B2 region has very extended blueshifted and redshifted lobes
with wide opening angles.
This outflow is most massive and have the largest momentum
among the identified outflows in the CO ($J=1-0$) map. 
We estimate the total energy injection rate due to the molecular 
outflows identified by the present and previous studies
to be about 0.2 $L_\odot$, larger than or at least comparable to 
the turbulence dissipation rate [$\approx (0.03 - 0.1) L_\odot$].
Therefore, we conclude that 
the protostellar outflows are likely to play a significant role in 
replenishing the supersonic turbulence in this clump.
\end{abstract}

\section{Introduction}
\label{intro}

It is now well accepted that the majority of stars
in our Galaxy form in clusters \citep{lada03}.  
Recent observations have revealed that active cluster forming regions 
are not distributed uniformly in parent molecular clouds, 
but localized and embedded in dense clumps with typical 
sizes of $\sim$ 1 pc and masses of $10^2 \sim$ 10$^3 M_\odot$
\citep[e.g.,][]{ridge03}.
More than half of all the young stellar populations associated 
with a parent molecular cloud are confined in such pc-scale, 
cluster forming clumps 
\citep[e.g.,][]{lada91,carpenter00,allen06}, where
many stars at different stages 
of formation are born in close proximity to one another. 
For example, in the Perseus molecular cloud, one of the nearby 
well-studied star forming regions, two pc-scale cluster forming clumps,
IC 348 and NGC 1333, contain about 80\% of the young 
stars associated with this cloud \citep{carpenter00}.
The mass fraction of molecular gas occupied by these two regions in the parent
molecular cloud is less than a few tens \%.
In such dense clumps, the feedback from the 
older generation of stars affects the formation of 
the younger generation of stars
\citep[e.g.,][]{norman80,mckee89,vazquez-semadeni10}.

Among the stellar feedback processes, 
protostellar outflows have been considered to be 
one of the important mechanisms to control the structure 
and dynamical properties of cluster forming clumps 
\citep{bally96,shu00,matzner00,matzner07}
because the outflows from a group of young stars interact with 
a substantial volume of their parent clump by sweeping up 
the gas into shells.
Indeed, recent numerical simulations of cluster formation
have demonstrated that the protostellar outflows largely regulate 
the structure formation and star formation in a dense cluster forming clump
\citep{li06,nakamura07,wang10,li10}. 
\citet{li06} showed that the supersonic turbulence in dense clumps can
be maintained by the momentum injection from the protostellar
outflows \citep[see also][]{carroll09}, 
and thus the clumps as a whole can be supported by
the turbulent  pressure due to the protostellar outflows
against global gravitational collapse 
at least for several dynamical times.
Furthermore, \citet{nakamura07} showed that the global star formation 
efficiency tends to be reduced by the momentum injection from the protostellar
outflows, although local star formation can often 
be triggered by the dynamical compression due to the protostellar outflows.
These studies indicate that the protostellar outflow feedback is a key 
to understanding the formation of stars in a pc-scale cluster forming clump.

Recent millimeter and submillimeter observations of nearby
pc-scale cluster forming clumps have revealed 
the significant role of protostellar outflows in clustered star
formation.  An excellent example is NGC 1333, where 
several protostellar outflows apparently overlapping and 
interacting with themselves are found in the central region of the
 dense clump; those outflows influence the distribution 
of the dense gas significantly \citep{sandell01,gutermuth09}.
The less dense parts, seen as cavities, are found to be 
filled with high-velocity gas created by the outflows
\citep{knee00,walsh07,hatchell07,hatchell09}.
On the basis of the $^{13}$CO ($J=1-0$) observations, 
\citet{quillen05} found 22 cavities with sizes of 0.1 $-$ 0.2 pc
in diameter in the central region of NGC 1333
\citep[see also][]{cunningham06}. 
Those cavities appear to expand slowly. 
They interpreted those cavities as remnants of past YSO outflow 
activity, although the role of those ``fossil'' cavities in cloud dynamics
is still unclear.

Another example is the Serpens clump, where several powerful
outflows were identified by the CO ($J=2-1$) observations
\citep{davis99}.
The energy injection rate of these outflows is comparable to or 
somewhat larger than the dissipation rate of the turbulent energy 
in this clump, indicating that the outflows are the main source 
of the supersonic turbulence \citep{sugitani10}.
\citet{sugitani10} suggested that the outflows in this clump
 have enough momentum to support the whole clump 
against global collapse.
For NGC 2264-C, \citet{maury09} suggested that
the parent clump cannot be sufficiently supported 
by the outflows in the current generation, 
although they are likely to contribute much to maintaining 
the observed turbulence.
In the Orion Molecular Cloud-2 FIR 3/4 region, \citet{shimajiri08} 
found an observational example showing that the dynamical interaction
of protostellar outflows with a dense clump 
may have triggered fragmentation of the clump,
leading to future star formation.
In this paper, to gain further evidence of the significant role 
of protostellar outflows in clustered star formation, 
we investigate the outflow activity in the 
dense part of the $\rho$ Ophiuchi molecular cloud 
(hereafter the $\rho$ Ophiuchi main cloud), 
one of the nearby pc-scale cluster forming clumps, on the basis of  
CO ($J=3-2$) and CO ($J=1-0$) observations.

The $\rho$ Ophiuchi main cloud is a suitable object  
to investigate the star formation process in a cluster forming clump
because this cloud is the nearest pc-scale cluster forming clump 
at a distance of about 125 pc (e.g., Lombardi et al. 2008; 
Loinard et al. 2008; Wilking et al. 2008). 
The $\rho$ Oph clump is known to harbor a rich cluster 
of young stellar objects (YSOs) in different evolutionary stages 
(Wilking et al. 2008; Enoch et al. 2008; J\o rgensen et al. 2008). 
On the basis of the recent infrared observations using the Spitzer 
space telescope, \citet{padgett08} identified about 300 YSOs, 
including Class 0/I/II/III sources, most of which 
are associated with the central region of the $\rho$ Ophiuchi main cloud.
Furthermore, recent millimeter and submillimeter dust continuum 
observations of $\rho$ Oph have revealed that a large number of dense cores 
are concentrated in the central dense part of the clump 
and their mass spectra are similar in shape to the stellar IMF 
\citep{motte98,johnstone00,stanke06}. 
These facts indicate that active star formation is still ongoing 
in this cloud.

So far, more than 10 molecular outflows from the YSOs have 
been identified in $\rho$ Oph 
\citep[e.g.,][]{bontemps96,sekimoto97,kamazaki03,wilking08}.  
The most spectacular outflow in $\rho$ Oph is the one from 
a prototypical Class 0 source, VLA 1623, \citep{andre90,dent95},
which is associated with Herbig-Haro objects, many knots of 
the H$_2$ emission at 2.12 $\mu$m, and the water masers.
\citet{dent95} mapped the molecular outflow from the VLA 1623 using  
CO$(J=3-2)$ emission. They revealed that the outflow is highly-collimated 
with an opening angle less than 1.6 degree, extending over a 
length of about 0.5 pc. \citet{kamazaki03} discovered three other
molecular outflows in the Oph A and B2 regions, although they did not 
successfully map the entire outflow lobes.  
In the southern part of $\rho$ Oph, \citet{bussmann07}
mapped the outflows from EL 29 and LFAM 26.
These outflows appear to be less powerful than the VLA 1623
outflow.  However, the physical properties of the
molecular outflows previously identified have not been well determined  
because of the lack of the large-scale mapping observations.  
To study how the outflows affect the clump dynamics and local star
formation, we have performed the CO ($J=3-2$) and CO ($J=1-0$) 
observations toward the $\rho$ Oph main cloud, using 
the ASTE 10 m and Nobeyama 45 m telescopes, respectively, 
and have derived the physical
properties of the outflows identified in the clump.

The rest of the paper is organized as follows. First, we
present the details of our CO $(J=3-2)$ and CO $(J=1-0)$
mapping observations in Section \ref{sec:observations}.
In Section 3, we derive the physical parameters of 
the molecular outflows identified from our CO ($J=3-2$) and 
CO ($J=1-0$) maps.
Then, we make a brief discussion of how the observed outflows 
interact with the dense cores identified from the 
H$^{13}$CO$^+$ ($J=1-0$) map by \citet{maruta10} 
in Section \ref{sec:discussion}.
We also estimate how the observed outflows contribute 
to turbulence generation in this clump, and 
compare the clump with other nearby cluster forming clumps, 
Serpens and NGC 1333.
Finally, we summarize our main conclusion in Section \ref{sec:summary}.

\section{Observations}
\label{sec:observations}

\subsection{CO ($J=3-2$) Observations}
\label{subsec:co32observations}

The CO ($J=3-2$; 345.7959899GHz) mapping observations were carried out 
with the ASTE 10 m telescope \citep{ezawa04}
during the period of 2004 October. 
The beamsize in HPBW of the ASTE telescope is 22$''$, which corresponds 
to 0.013 pc at a distance of 125 pc.
The main-beam efficiency, $\eta_{\rm 32}$, was 
0.35 at 345 GHz.
We used a 345 GHz SIS heterodyne receiver, which had the 
typical system noise temperature of 175-350 K in DSB mode 
at the observed elevation. 
The temperature scale was determined by the chopper-wheel method, 
which provides us with the antenna temperature, $T_A^*$, corrected 
for the atmospheric attenuation. As a backend, we used four sets 
of 1024 channel autocorrelators, providing us with a frequency 
resolution of 125 kHz that corresponds to 0.11 km s$^{-1}$
at 345 GHz. 
The position switching technique was employed to cover 
the dense region of the $\rho$ Ophiuchi main cloud, 
about $23'\times 23'$ area,
corresponding to about 0.8 pc $\times$ 0.8 pc at a distance of 125 pc. 
After subtracting linear baselines, the data were convolved with a 
Gaussian-tapered Bessel function (Magnum et al. 2007) 
and were resampled onto a 5$''$ grid. 
The resultant effective FWHM resolution is 40$''$.
The typical rms noise level is 0.28 K in brightness temperature
($T_{\rm mb}$)
with a velocity resolution of 0.4 km s$^{-1}$.

\subsection{CO ($J=1-0$) Observations}
\label{subsec:co10observations}

As shown in the next section, our CO ($J=3-2$) observations
revealed that the outflow previously identified by \citet{kamazaki03} 
in the Oph B2 region, the EL 32 outflow, is very extended.
Unfortunately, the CO ($J=3-2$) map does not cover the whole 
redshifted outflow lobe. To map the whole extent of the EL 32 outflow, 
we performed the CO ($J=1-0$) observations using 
the Nobeyama 45 m radio telescope.
The CO ($J=1-0$; 115.2712018 GHz) mapping observations were carried out 
with the 25-element focal plane receiver BEARS equipped in the 
Nobeyama 45 m telescope during the period from December 2009 to May
2010, as a part of the Nobeyama 45m Legacy Project 
(see {http://www.nro.nao.ac.jp/}). 
At 115 GHz, the telescope has a FWHM beam size of 15$''$ and a main beam
efficiency, $\eta _{10}$, of 0.32.
At the back end, we used 25 sets of 1024 channel auto-correlators (ACs)
which have bandwidths of 32 MHz and frequency resolutions of 37.8
kHz \citep{sorai00}.
The frequency resolution corresponds to a velocity resolution of 
0.1 km s$^{-1}$ at 115 GHz.
During the observations, the system noise temperatures were in the
range between 300 to 900 K in DSB at the observed elevation.
The standard chopper wheel method was used to convert the output signal
into  the antenna temperatures ($T_{A}^*$), corrected for the
atmospheric attenuation.
Our mapping observations were made by the OTF mapping technique 
\citep{sawada08}. 
We adopted a spheroidal function as a gridding convolution function (GCF)
to calculate the intensity at each grid point of the final cube
data with a spatial grid size of 12$''$. 
The final effective resolution of the map is 30$''$.
The rms noise level of the final map is 1.0 K
in brightness temperature ($T_{\rm mb}$) at a
velocity resolution of 0.4 km s$^{-1}$.

\section{Results}

\subsection{CO ($J=3-2$) and CO ($J=1-0$) maps}
\label{subsec:co32+co10}

\begin{figure}
\epsscale{1.0}
\plotone{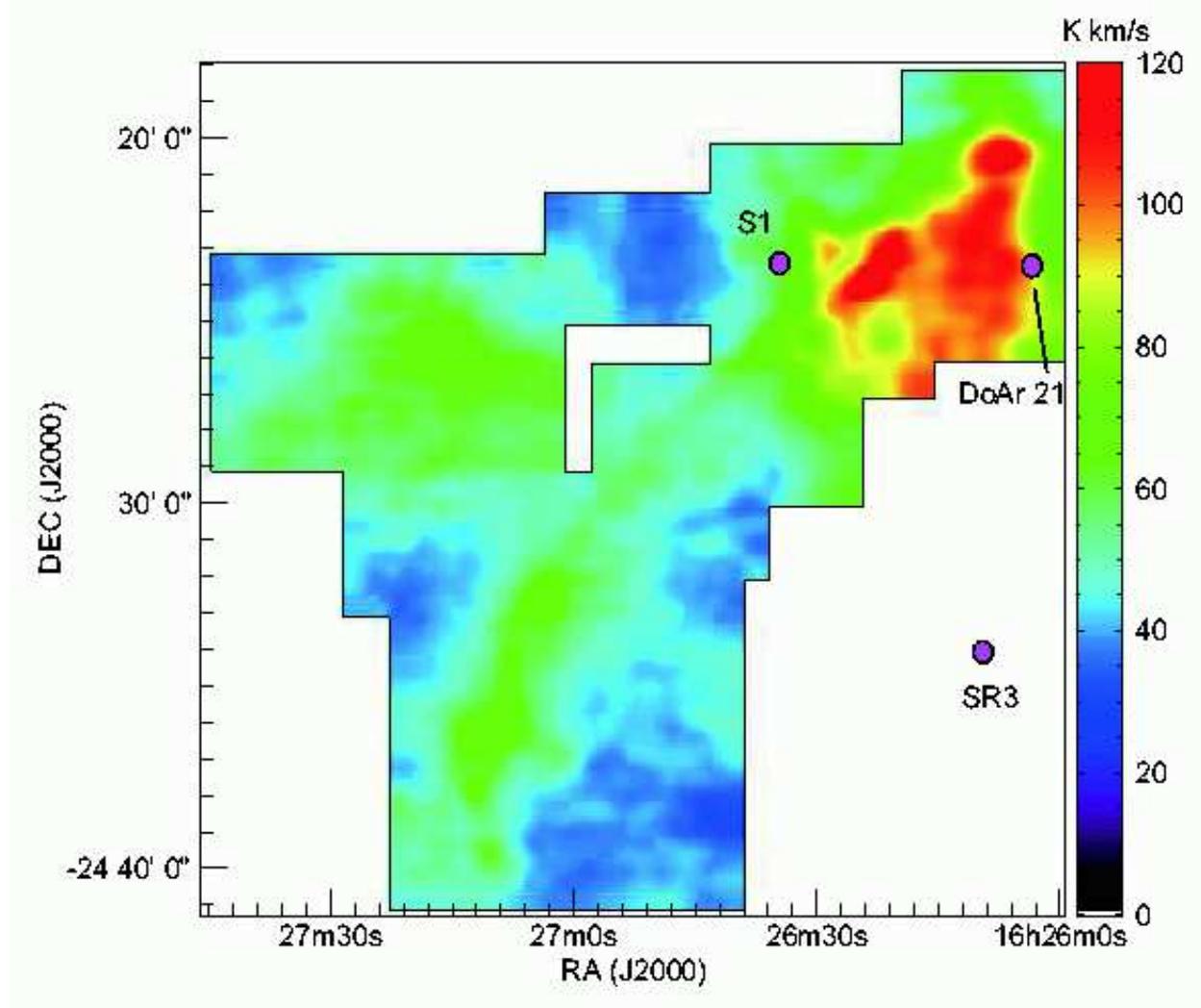}
\caption{
CO ($J=3-2$) total integrated  intensity map in the velocity
range from $v_{\rm LSR} = -9.8$ km s$^{-1}$ to +12.2 km s$^{-1}$
toward the $\rho$ Ophiuchi main cloud.
The CO ($J=3-2$) integrated intensity is shown 
in color in units of K km s$^{-1}$.
The circles indicate the positions of 
 B stars S1, DoAr 21, and SR3.
}  
\label{fig:integ}
\end{figure}

\begin{figure}
\epsscale{1.0}
\plotone{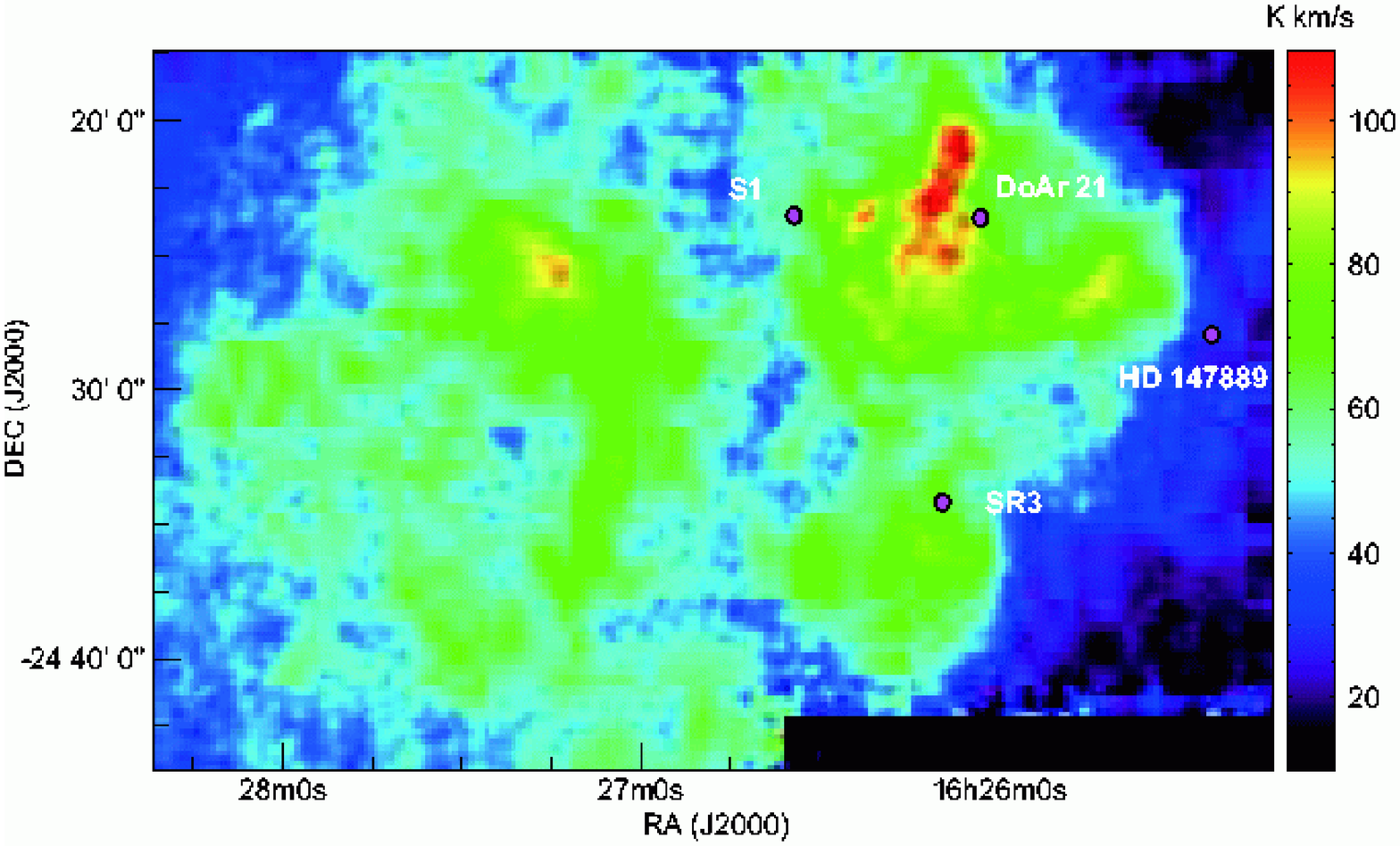}
\caption{
CO ($J=1-0$) total integrated  intensity map in the velocity
range from $v_{\rm LSR} = -3.2$ km s$^{-1}$ to +8.0 km s$^{-1}$
toward the $\rho$ Ophiuchi main cloud.
The CO ($J=1-0$) integrated intensity is shown in color 
in units of K km s$^{-1}$.
The circles indicate the positions of 
 B stars S1, SR3, DoAr 21, and HD 147889.
}  
\label{fig:integ2}
\end{figure}

\begin{figure}
\epsscale{0.6}
\plotone{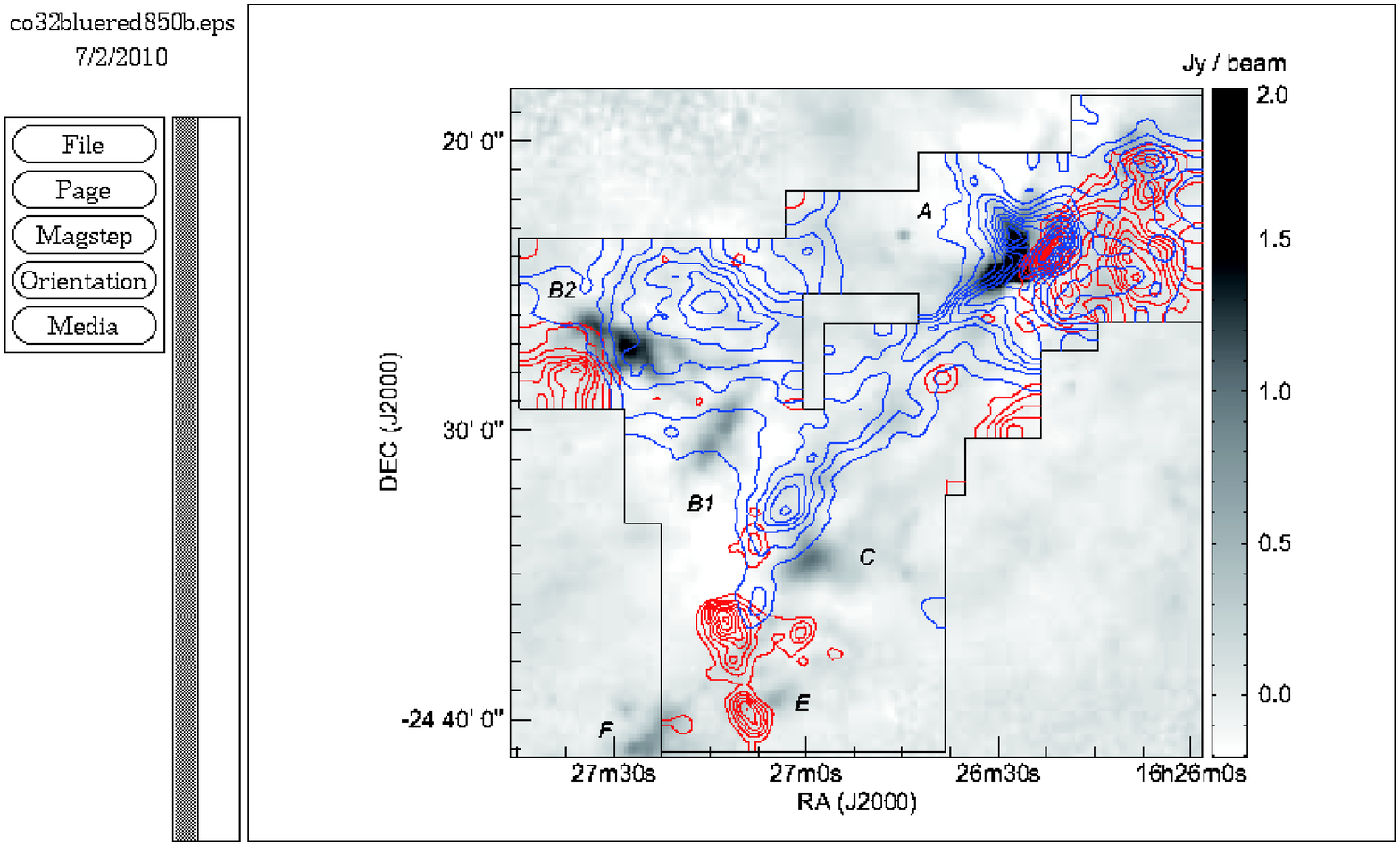}
\plotone{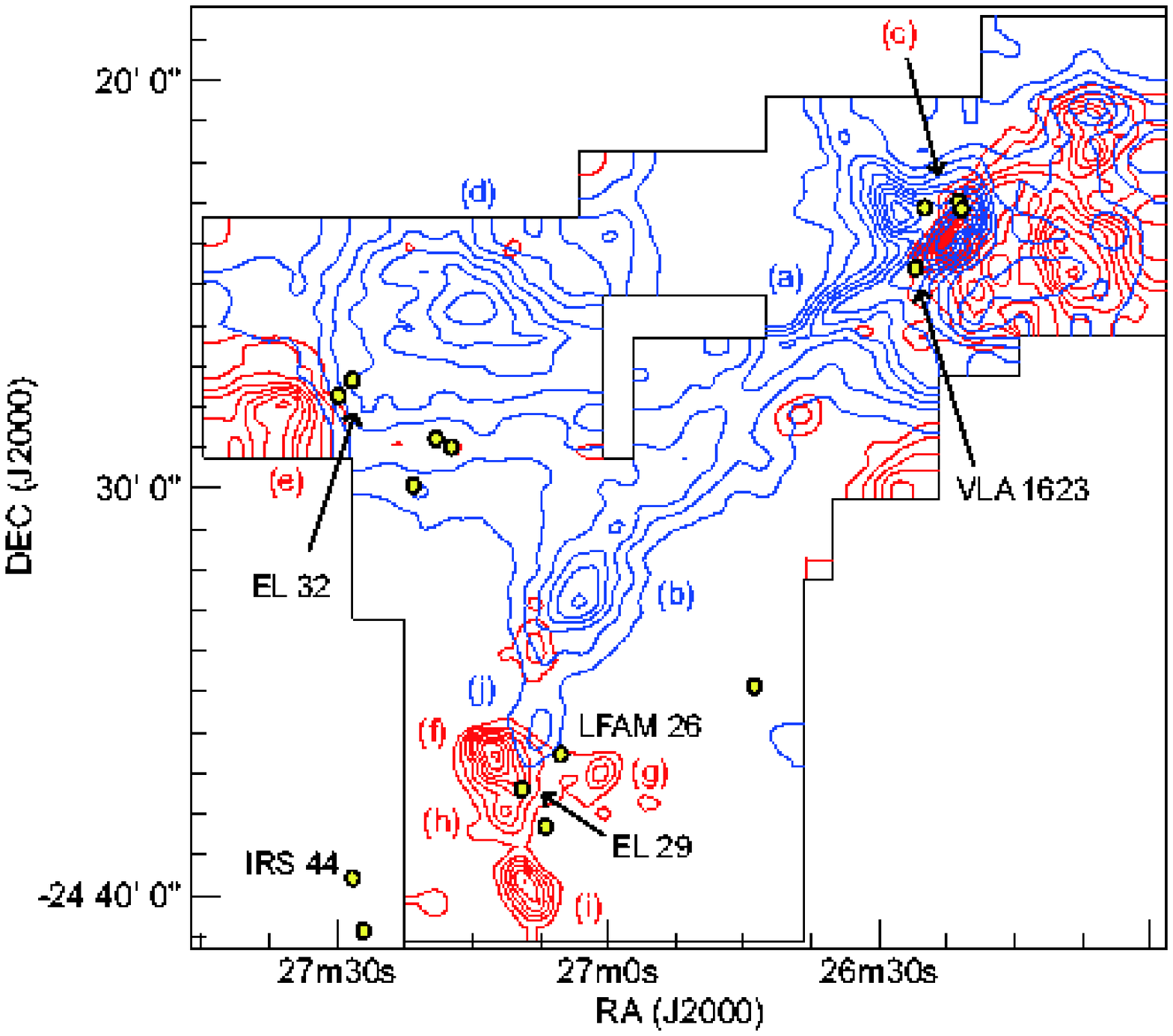}
\caption{
(a) CO ($J=3-2$) map of the $\rho$ Ophiuchi main cloud
on the 850 $\mu$m image obtained by \citet{johnstone00}.
Blue contours represent blueshifted CO ($J=3-2$) intensity 
integrated between $-9.8$ km s$^{-1}$ and 2.2 km s$^{-1}$, 
starting from 18 K km s$^{-1}$ at intervals of 4 K km s$^{-1}$. 
Red contours represent redshifted CO ($J=3-2$) intensity 
integrated between $7.4$ km s$^{-1}$ and 11.8 km s$^{-1}$, 
starting from 1.6 K km s$^{-1}$ at intervals of 0.8 K km s$^{-1}$. 
The gray scale shows the 850 $\mu$m image in units of $Jy/{\rm beam}$.
The dense subclumps are designated by A, b1, B2, C, E, and F.
(b) CO ($J=3-2$) map of the $\rho$ Ophiuchi main cloud
overlaid with the positions of the embedded YSOs 
listed by \citet{vankempen09} and some other YSOs with the yellow circles.
Positions of several peaks of outflow components 
 detected by the CO ($J=3-2$) observations are indicated
by the alphabets (a) through (j).
}  
\label{fig:map1}
\end{figure}

In Figures \ref{fig:integ} and \ref{fig:integ2}, 
we present the CO ($J=3-2$) and CO ($J=1-0$) integrated intensity maps 
in the ranges of $v_{\rm LSR}=-9.8 \sim +12.2$ km s$^{-1}$
and $v_{\rm LSR}=-3.2 \sim +8.0$ km s$^{-1}$, respectively.
In Figures \ref{fig:map1} and \ref{fig:map2}, 
the distributions of the blueshifted and redshifted 
CO ($J=3-2$) and CO ($J=1-0$) components, respectively,  
are also shown on the 850$\mu$m images obtained by \citet{johnstone00}.  
In Figure \ref{fig:map1}a, 
several dense subclumps, which are recognized in the 850 $\mu$m images, 
are designated by A, B1, B2, C, D, E, and F.
By comparing the H$^{13}$CO$^+$ spectra with
the $^{12}$CO ($J=3-2$ and $J=1-0$) ones at the same position, 
the shapes of the $^{12}$CO ($J=3-2$) and $^{12}$CO ($J=1-0$)
line profiles appear to be caused mainly by self-absorption 
rather than multiple velocity components (see also Figures 
\ref{fig:profile} and \ref{fig:profile2}).

The CO $(J=1-0)$ map in Figure \ref{fig:integ2} shows a clear break 
between the east and west parts of the $\rho$ Oph clump, although
the radial velocities of the two parts are not so different
($v_{\rm LSR} \sim 3.5 - 4.5$ km s$^{-1}$).
The CO $(J=3-2)$ integrated intensity map also appears to 
reasonably follow that of CO $(J=1-0)$, although the 
detailed distributions are different.
The strongest CO ($J=3-2$) emission comes from the western side of Oph A
(see Figure \ref{fig:integ}). 
As shown later, the region with the strongest CO ($J=3-2$) emission 
has both the blueshifted and redshifted components that 
are likely to be due to the outflow from the VLA 1623 source.
In the entire Oph A filament,
the CO ($J=3-2$) component blueshifted from the systemic 
velocity of Oph A is predominantly strong, 
suggesting that Oph A is compressed from the far side by the 
photodissociation region (PDR) excited by a young B3 star, S1.

The second strongest CO ($J=3-2$) emission comes near the north-west
edge of the observational box, in which
the CO ($J=1-0$) integrated intensity is the strongest.  
This region has both broadly distributed blueshifted and redshifted 
components, and may be influenced 
by the PDR excited by the B2 star, HD 147889.
There are two other areas having moderately strong CO ($J=3-2$) emission.
One is located near the north-east part of the CO ($J=3-2$)
observational box, and 
the other is an elongated structure running from the box center 
to the south-east part.  
As discussed below, these two components originate from the high 
velocity gas by molecular outflows.

\subsection{Outflow Identification}

Using the CO ($J=3-2$) and CO ($J=1-0$) data cubes, 
we identified molecular outflows as follows 
\citep[e.g.,][]{takahashi08}.
First, we scrutinized the velocity channel maps of both the 
data to find localized blueshifted or redshifted emission 
in the area within a radius of 1$'$ from  each YSO.
Here, we used the YSO list obtained by 
\citet{jorgensen08} and \citet{vankempen09}.
From our CO ($J=3-2$) data, we identified 4 outflows.
All these outflows can also be clearly recognized 
in the CO $(J=1-0)$ map.
In addition, we identified one outflow from the 
CO ($J=1-0$) data, which is located outside 
our CO ($J=3-2$) map.
All the outflows identified in the present paper were 
already detected by previous studies.  
The characteristics of the identified outflows 
are summarized in Table \ref{tab:outflow}.
In Figures \ref{fig:profile} and \ref{fig:profile2}, 
we present the CO ($J=3-2$) and CO ($J=1-0$) spectra
centered on some of the high-velocity lobes, respectively.
Each of the spectra shows signs of significant self-absorption due to 
the ambient gas at ambient velocities; both the CO line emission is
clearly optically thick. 
In the following, we discuss the characteristics of the outflows
identified in the present paper, and then derive 
their physical parameters.

\subsection{Individual Outflows}

\subsubsection{VLA 1623}

In Figure \ref{fig:map1}, two highly-collimated, 
strong blueshifted-components labeled (a) and (b) 
can be clearly recognized.
They trace the collimated (south-east) outflow 
from the VLA 1623 source, which was first discovered 
by \citet{andre90} with the CO ($J=2-1$) observations. 
The north-west lobe of this outflow, labeled (c), 
is less prominent than the south-east one.
The north-west flow has the low-velocity blue and redshifted 
components, suggesting that the outflow axis 
is almost parallel to the plane of the sky.
This is in good agreement with the CO ($J=2-1$) observations 
of \citet{andre90}.
However, as shown in the next subsection, some physical quantities 
of the outflow are significantly different from theirs.
This is due to the fact that their observations did not cover 
the whole VLA 1623 outflow. 
For example, our outflow size is larger by a factor of 5,
the outflow velocity is smaller by a factor of 2, 
and thus  the estimated dynamical time 
of the south-east component is longer 
by a factor of about 10 than that estimated 
by \citet{andre90}.
When the outflow parameters are estimated using 
the same area as they mapped, our quantities agree 
reasonably with those derived by \citet{andre90}.

The whole outflow lobes [the components (a), (b), and (c)] were 
first mapped by \citet{dent95} using the CO $ (J=3-2)$ observations.
They found that the blueshifted flow is extended over a 
length of about 0.5 pc with a narrow opening angle 
of less than 1.6 degree.  Our CO ($J=3-2$) map is in good agreement 
with that of \citet{dent95}.  
Since this component is distinct in our CO $(J=1-0)$ map
presented in Figure \ref{fig:map2}, we think that 
it can be detected with the CO ($J=2-1$) emission.
Our total mass and momentum of the outflow are 
reasonably in good agreement with those of \citet{dent95},  
taking into account the different adopted fractional abundance ratio 
and distance to $\rho$ Oph, which cause the difference
in the mass estimation by a factor of 3.
The blueshifted components are also remarkable 
in the line profiles presented in Figures \ref{fig:profile} 
and \ref{fig:profile2}, 
in which the line profiles of the components (a) and (b) 
show strong blueshifted wings.

In Figure \ref{fig:h2knots}, we show the positions of the H$_2$ knots 
identified by \citet{gomez03} and \citet{khanzadyan04}. 
Many H$_2$ knots are associated with the blueshifted lobe labeled (a)
in Figure \ref{fig:map1}b.  
These H$_2$ knots are clearly recognized in the IRAC 
channel 2 (4.5$\mu$m) image taken by the Spitzer space 
telescope \citep{zhang10}.
\citet{khanzadyan04} compared their H$_2$ knots with those identified
by \citet{gomez03} to measure the proper motions of the H$_2$ knots, and 
estimated the flow speed of about 130 km s$^{-1}$, which is much faster 
than the outflow speed of about 20 km s$^{-1}$
estimated from our CO ($J=3-2$) data.

\subsubsection{Elias 32 and IRS 44}

\begin{figure}
\epsscale{1.0}
\plotone{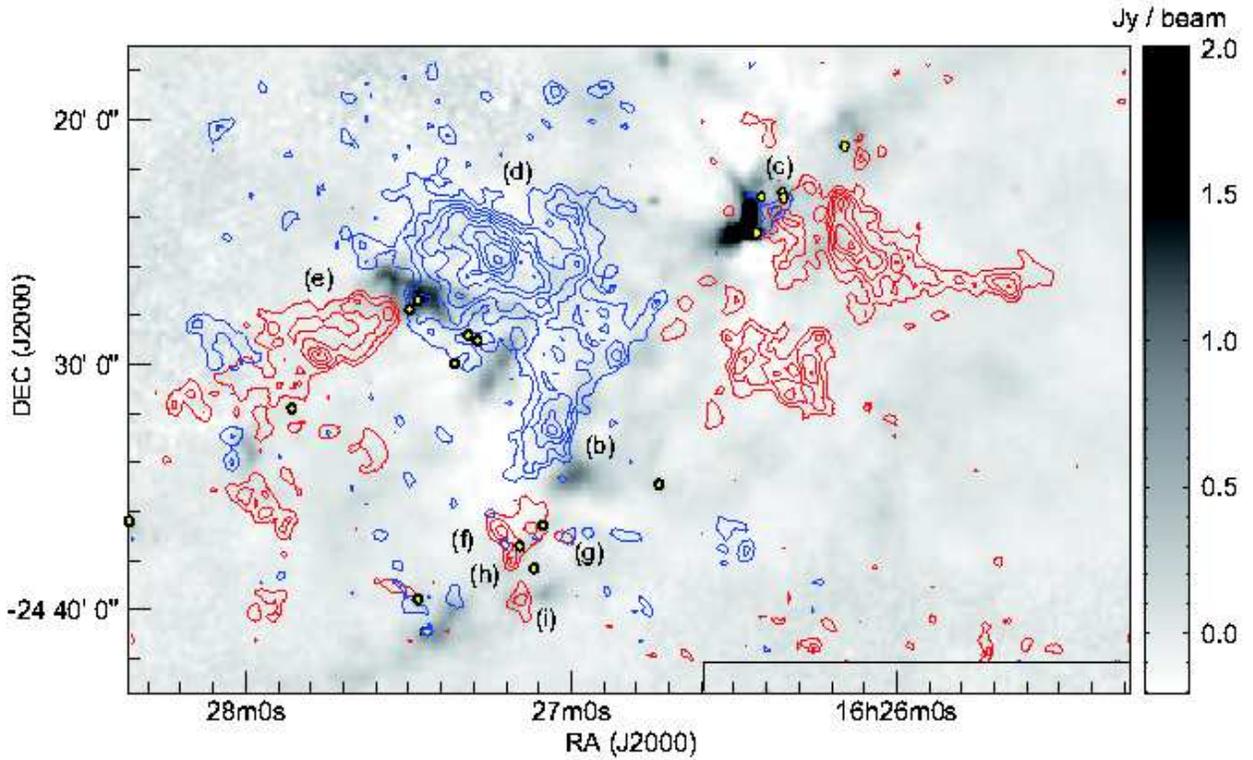}
\caption{
CO ($J=1-0$) map of the $\rho$ Ophiuchi main cloud
on the 850 $\mu$m image obtained by \citet{johnstone00}.
Blue contours show the CO ($J=1-0$) intensity integrated between 
$-4.0$ km s$^{-1}$ and 1.2 km s$^{-1}$, starting from 8.9 K km s$^{-1}$
at intervals of4.1 K km s$^{-1}$. 
Red contours show the CO ($J=1-0$) intensity integrated between 
6.4 km s$^{-1}$ and 9.2 km s$^{-1}$, starting from 4.1 K km s$^{-1}$
at intervals of 2.8 K km s$^{-1}$. 
The gray scale, symbols, and alphabets are the same as 
those of Figure \ref{fig:map1}.
}  
\label{fig:map2}
\end{figure}

\begin{figure}
\epsscale{0.8}
\plotone{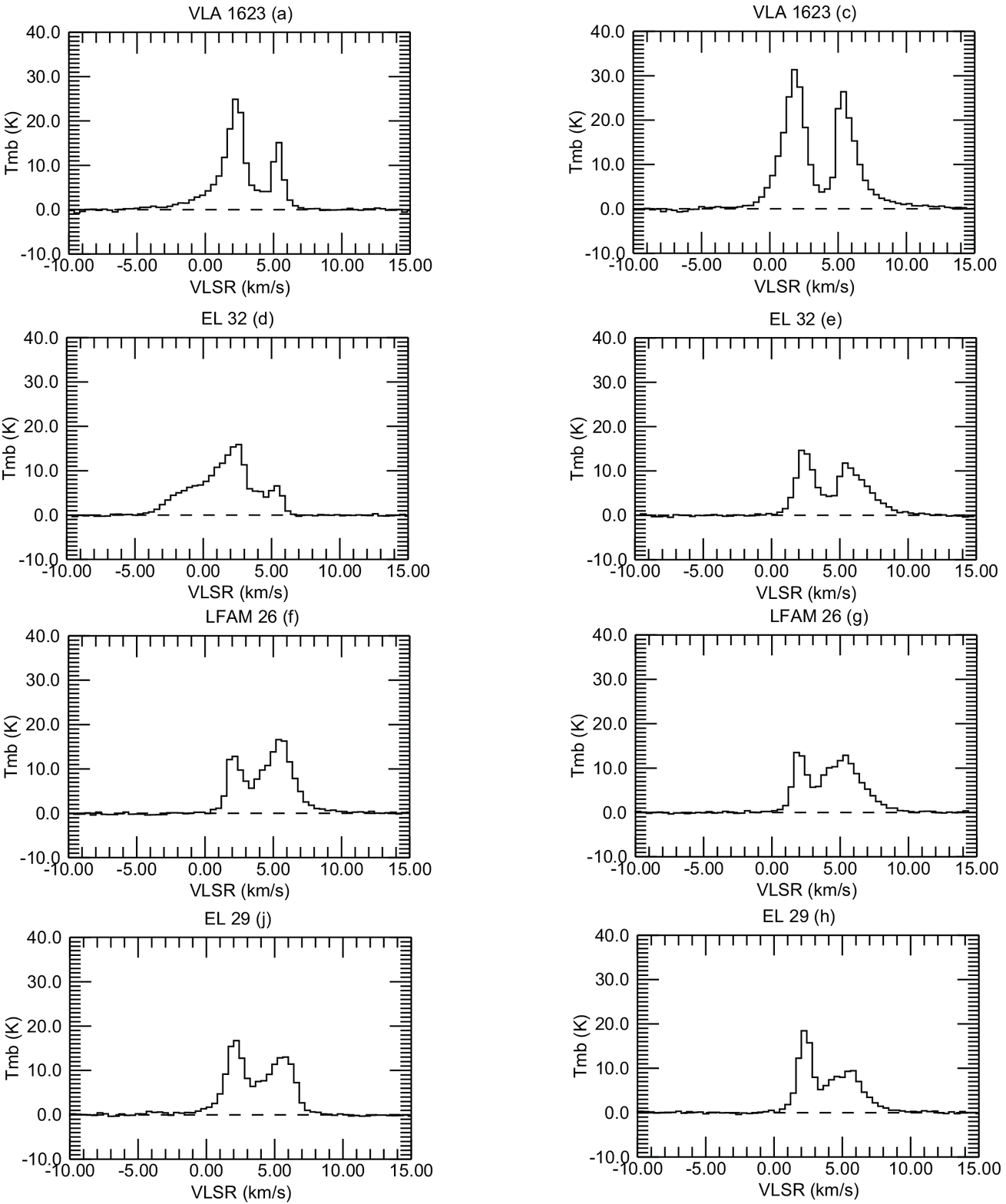}
\caption{
CO ($J=3-2$) spectra toward several high-velocity lobes 
indicated by alphabets in Figure \ref{fig:map1}. 
Each spectra is smoothed in a $1' \times 1'$ grid centered at the
position of the peak intensity.
}  
\label{fig:profile}
\end{figure}

\begin{figure}
\epsscale{0.8}
\plotone{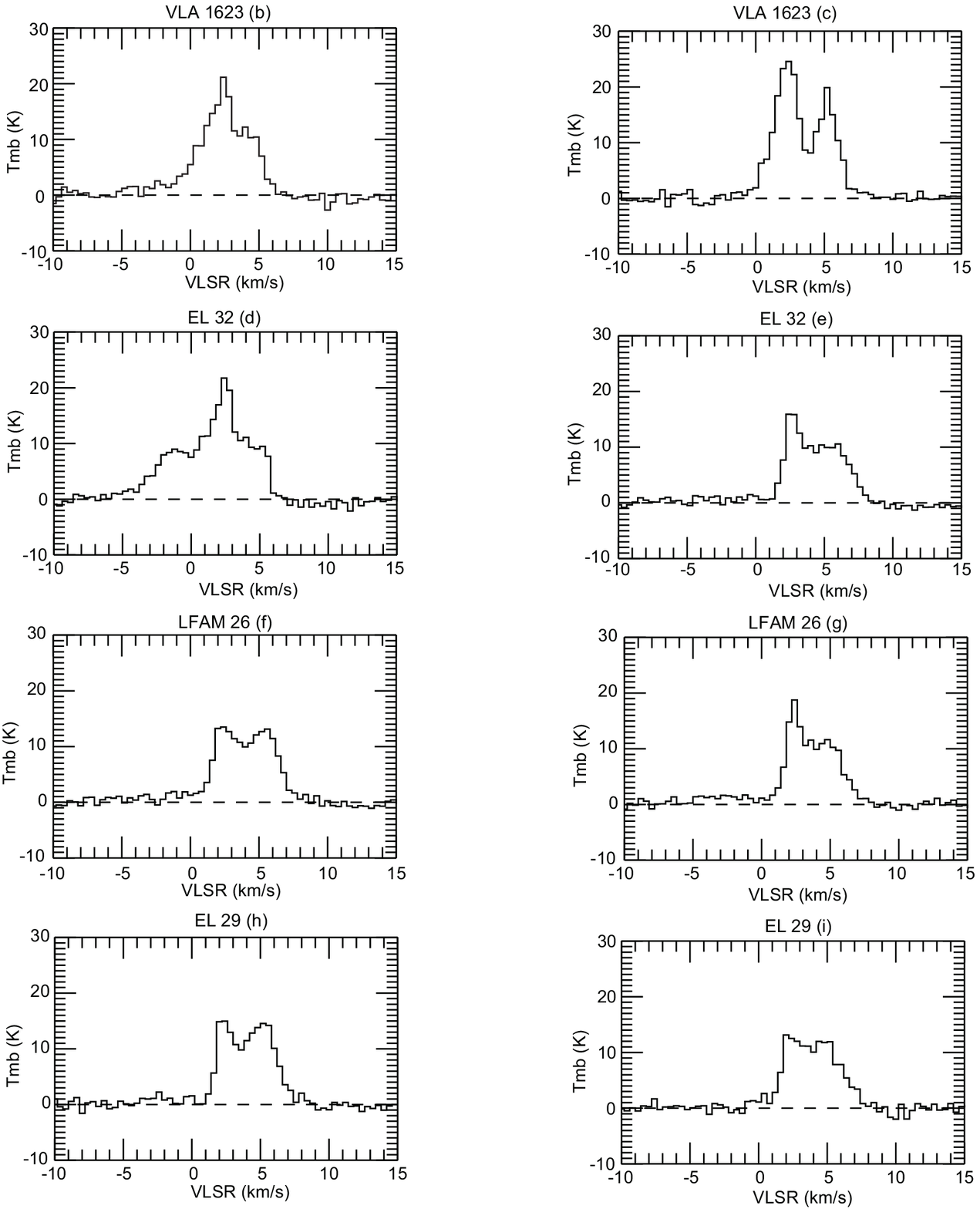}
\caption{
CO ($J=1-0$) spectra toward several high-velocity lobes 
indicated by alphabets in Figure \ref{fig:map2}. 
Each spectra is smoothed in a $1' \times 1'$ grid centered at the
position of the peak intensity.
}  
\label{fig:profile2}
\end{figure}

\begin{figure}
\epsscale{1.0}
\plotone{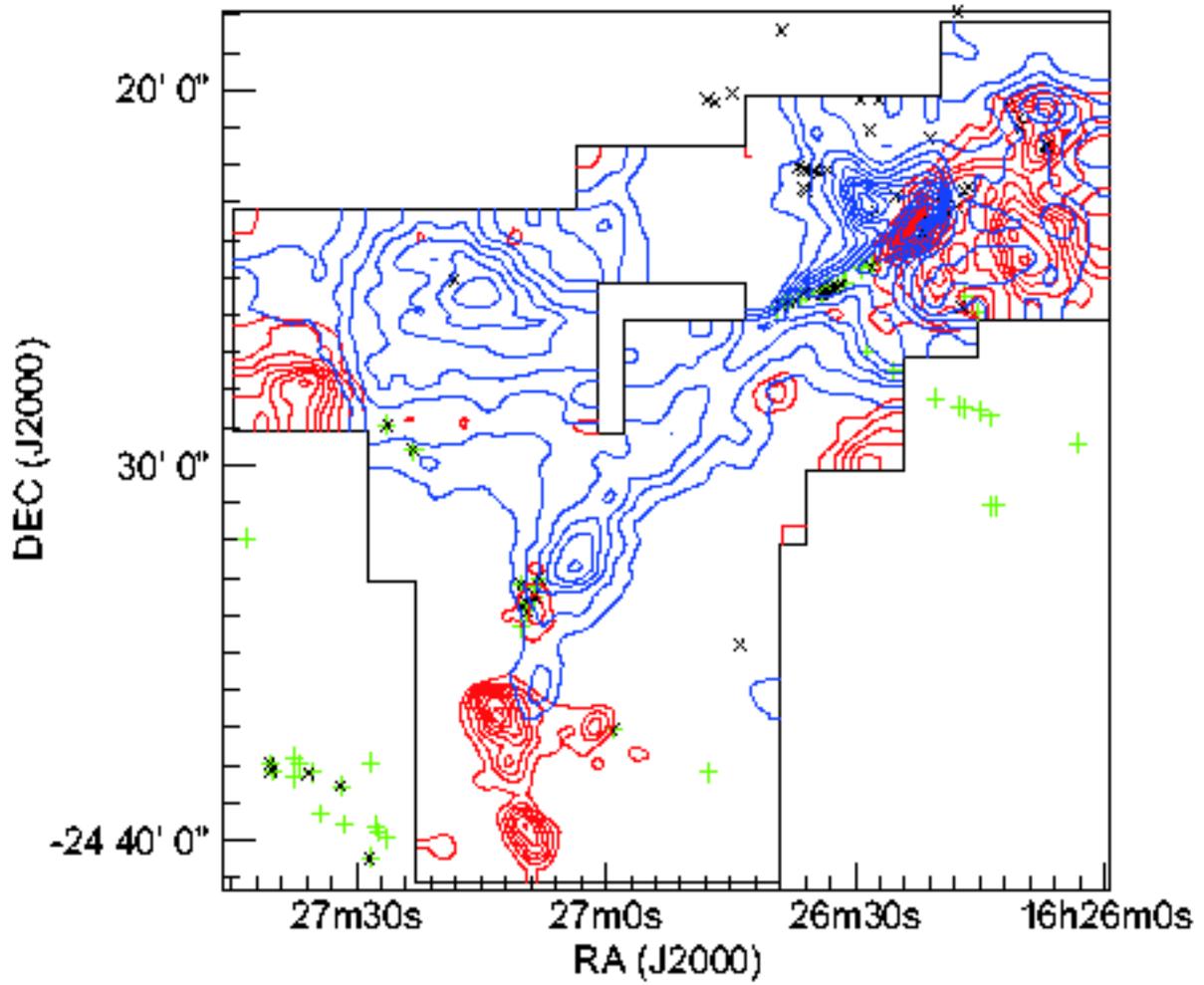}
\caption{CO ($J=3-2$) map of the $\rho$ Ophiuchi main cloud.
The contours are the same as those of Figure \ref{fig:map1}.
The crosses and pluses indicate the positions of 
the H$_2$ knots identified by \citet{gomez03} and 
\citet{khanzadyan04}, respectively.
}  
\label{fig:h2knots}
\end{figure}

In our CO ($J=3-2$) image shown in Figure \ref{fig:map1}, we can 
clearly recognize a spatially extended outflow in the Oph B2 region
(labeled (d) and (e)).
The line profiles presented in Figures \ref{fig:profile} and 
\ref{fig:profile2} also show distinct high-velocity wing emission  
for both the blueshifted and redshifted components.
In particular, the blueshifted wings of both the 
CO ($J=3-2$) and CO ($J=1-0$) line profiles are very strong.
This outflow appears to be from a Class I YSO, Elias 32, as suggested 
by \citet{kamazaki03}.
Although in our CO ($J=3-2$) image, most of the redshifted component 
labeled (e)
appears to be outside the observed area, our CO (J$=1-0$)
map have successfully revealed that the redshifted flow 
also extends with a wide opening angle 
toward the south-east direction (see Figure \ref{fig:map2}).

The blueshifted (north-west) lobe appears to be anti-correlated 
with the dense gas detected in H$^{13}$CO$^+$ emission 
in the B2 region by \citet{maruta10} (see Figure \ref{fig:map3}a).
Around the peak of the CO ($J=3-2$) emission in the 
blueshifted lobe, the H$^{13}$CO$^+$ emission is less prominent, indicating 
that the dense gas is swept out of the region.
In addition, the dense gas well follows the south-east part 
of the blueshifted lobe, showing an arc-like shape.
According to \citet{maruta10}, a small hole is seen
in their H$^{13}$CO$^+$ data cube 
in the southern part of the Oph B2 region 
(labeled No. 1 in their Figures 2b and 2c),
which is located near the north-west edge of 
the redshifted lobe.
Two other holes are also seen near the northern (No. 2
in their Figure 2b, corresponding to the south-east edge
of the blueshifted lobe) and southern (No. 3 in their Figure 2b,
corresponding to the north-west edge of the redshifted lobe) 
edges of the B2 region. 
These holes are likely to be created by the EL 32 outflow.
Furthermore, the blueshifted lobe is extended beyond the B2 region,
reaching the No. 4 and 5 arcs in Figure 2b of \citet{maruta10}. 
These structures suggest that the Oph B2 region is strongly 
influenced by the EL 32 outflow.
See Section 3 of \citet{maruta10} for more detail.

Near the southern boundary of the observed area with 
CO ($J=1-0$) presented in Figure \ref{fig:map2},
which is outside the  CO ($J=3-2$) map, we can see a weak blueshifted 
and redshifted lobes, which are probably the outflow lobes from IRS 44.
This outflow was first discovered by \citet{terebey89} by means of the
CO ($J=1-0$) interferometric observations 
using the Owens Valley Millimeter Interferometer.
\citet{jorgensen09} presented the three-color Spitzer IRAC image
of several knots that are proposed to be associated with the 
outflow from IRS 43 or other nearby YSO including IRS 44 
(see their Figure B.1). 
Although these knots are clearly along a line pointing directly
back toward both IRS 43 and 44, the direction of our redshifted lobe 
of the IRS 44 outflow does not match that suggested by the knots.
Thus, these knots are probably not associated with the IRS 44 outflow.

\subsubsection{Elias 29 and LFAM 26}

In the southern part of the CO ($J=3-2$) map (Figure \ref{fig:map1}), we 
can find  4 redshifted (labeled (f), (g), (h), and (i)) 
and 1 blueshifted (labeled (j)) components.
This region was mapped by \citet{bussmann07} using 
the CO ($J=3-2$) emission, but our map covers a larger
area than that of \citet{bussmann07}.
The overall feature of the CO ($J=3-2$) map is in good agreement with 
that of \citet{bussmann07}, although we found a new redshifted
component labeled (i) in the southern part of this region.  
The redshifted components labeled (h) and (i)
 and the blueshifted component labeled (j) are most likely to be
from a Class I YSO, EL 29. 
As pointed out by \citet{bussmann07}, 
the EL 29 outflow axis has an S-like shape, 
traced by several H$_2$ knots detected by 
\citet{gomez03} and \citet{ybarra06}.
In addition, our CO ($J=3-2$) and CO ($J=1-0$) maps 
in Figures \ref{fig:map1} and \ref{fig:map2} show that the 
EL 29 outflow has both the blueshifted and redshifted components 
on the north side,  suggesting that the outflow axis is 
almost parallel to the plane of the sky. 
The reason why the EL 29 outflow is so deformed is unclear.
One possibility is the interaction with the magnetic field.
Recently, the magnetic field structure toward $\rho$ Oph 
has been mapped by \citet{tamura10}. They found that the 
ordered magnetic field penetrates the Oph E and F filaments
almost perpendicularly. The direction of the magnetic field
coincides well with the outflow axis of the redshifted component.
Another possibility is the dynamical interaction with dense gas.
In fact, the redshifted outflow lobe is located in the 
less dense region between the Oph E and F regions. This implies
that the outflow hit the dense gas ridge 
and the direction was altered.  Both effects might be responsible 
for the shape of the redshifted outflow lobe.

On the other hand, we interpret that the other redshifted components (f) 
and (g) are associated with the outflow from a Class I YSO, LFAM 26, 
because a line connecting the two H2 knots detected 
by \citet{khanzadyan04}, f04-01a and f04-01b, reasonably 
coincides with a line between the outflow components (f) and (g).
Furthermore, the bow-shocked-like shapes of the two knots support our
interpretation that the component (g) is coming from LFAM 26.
As seen in the CO ($J=1-0$) map in Figure \ref{fig:map2},
the redshifted components labeled (f) and (g) 
 are also associated with
weak blueshifted components, suggesting that the LFAM 26 outflow 
axis is nearly parallel to the plane of the sky.
This interpretation is consistent with the fact that 
the disk around LFAM 26 is almost edge-on 
with the rotation axis parallel to the east-west direction
\citep{duchene04}.
Although the VLA 1623 outflow is still the most powerful in this
region, the estimated speeds of the EL 29 and LFAM 26 outflows are 
relatively as large as that of VLA 1623 when the inclination 
angles of the three outflows are taken into account.

\subsection{Derivation of Physical Quantities}

In the above subsection, we identified 5 outflows 
from the CO ($J=3-2$) 
and CO ($J-1-0$) observations.
Here, we derive the physical quantities of the 
5 clear outflows, following the procedure described below.

Under the assumption of local thermodynamical equilibrium (LTE)
condition and optically thin emission, 
the molecular outflow masses derived from CO ($J=3-2$) and CO ($J=1-0$) 
can be calculated, respectively, as  
\begin{equation}
M_{\rm 32} = \sum _j M_{32,j}  \ ,
\end{equation}
and 
\begin{equation}
M_{\rm 10} = \sum _j M_{10,j} \ .
\end{equation}
Here, $M_{32,j}$ and $M_{10,j}$ are the outflow masses at the 
$j$-th channel, derived from CO ($J=3-2$) and CO ($J=1-0$),
respectively, and are given by 
\begin{eqnarray}
M_{32,j} &=&  5.0 \times 10^{-8}
\left({X_{\rm CO} \over 10^{-4}}\right)^{-1}
\left({D \over 125 {\rm pc}}\right)^2
\left({\Delta \theta \over {\rm arcsec}}\right)^2 \nonumber \\
&& \times \left({\eta_{32} \over 0.35}\right)^{-1}
\left({\sum _i T_{A,i,j}^* (3-2) \Delta v 
\over {\rm K \ km \ s^{-1}}}\right)  M_\odot
 \ ,
\end{eqnarray}
and 
\begin{eqnarray}
M_{10,j} &=& 3.4 \times 10^{-7}
\left({X_{\rm CO} \over 10^{-4}}\right)^{-1}
\left({D \over 125 {\rm pc}}\right)^2
\left({\Delta \theta \over {\rm arcsec}}\right)^2 \nonumber \\
&& \times \left({\eta_{10} \over 0.32}\right)^{-1}
\left({\sum _i T_{A,i,j}^*(1-0) \Delta v \over {\rm K \ km \ s^{-1}}}\right) 
 M_\odot
 \ ,
\end{eqnarray}
where $i$ denotes the grid index on the $j$-th channel.
Here, the same excitation temperature $T_{\rm ex}$ of 30 K
is adopted for the both lines.
The fractional abundance of CO relative to H$_2$, $X_{\rm CO}$,
and the distance to the cloud, $D$,  are adopted to be 
 $1\times 10^{-4}$ 
(a typical value of high-density regions of molecular clouds;
e.g., \citet{garden91}) and 
125 pc \citep[e.g.,][]{lombardi08,loinard08,wilking08}, 
respectively.
The main beam efficiencies of the ASTE and Nobeyama 45 m telescopes,
are $\eta_{\rm 32}=0.35$ and $\eta_{\rm 10}=0.32$, respectively.
The integrated intensities of the CO ($J=3-2$) and CO ($J=1-0$) lines
at the $i$-th grid are $\eta_{32}^{-1} \sum_j T_{A, i,j}^*(3-2) \Delta v$ and 
$\eta_{10}^{-1} \sum_j T_{A, i,j}^* (1-0) \Delta v$,
respectively, where we integrated the antenna temperatures above the 
3 $\sigma$ noise levels in the velocity range given 
in Tables \ref{tab:co32} and \ref{tab:co10}
(see Sections \ref{subsec:co32observations} and  \ref{subsec:co10observations}
for the values of the 1 $\sigma$ rms noise).
The symbols $\Delta \theta$ and $\Delta v $ (= 0.4 km s$^{-1}$) 
indicate the grid size and the velocity resolution of the data.

For both the lines, the outflow mass $M_{\rm out}$, 
the outflow momentum $P_{\rm out}$, and kinetic energy 
$E_{\rm out}$ are estimated as
\begin{equation}
M_{\rm out}=\sum_j M_j \ , 
\end{equation}
\begin{equation}
P_{\rm out}=\sum_j M_j |V_j - V_{\rm sys}| / \cos \xi  \ , 
\end{equation}
and
\begin{equation}
E_{\rm out} = \sum _j \frac{1}{2} M_j (V_j-V_{\rm sys})^2  / \cos^2 \xi \ ,
\end{equation}
respectively, where 
$M_j$ is the mass of the $j$-th channel,
$V_j$ is the LSR velocity of the $j$-th channel and 
$V_{\rm sys}$ is the systemic velocity of the driving source.
We adopt $V_{\rm sys}\simeq 3.8$, 4.3, 4.9, 4.9, and 4.2 km s$^{-1}$
for VLA 1623, EL 32, EL29, LFAM 26, and IRS 44, respectively. 
Each systemic velocity was determined from the peak $V_{\rm LSR}$ 
of the H$^{13}$CO$^+$ ($J=1-0$) emission averaged in
the $2'\times 2'$ area centered at the position of each source,
where we used the H$^{13}$CO$^+$ ($J=1-0$) emission of 
the Nobeyama 45 m archive data whose fits  data
are available from the Nobeyama web page (http://www.nro.nao.ac.jp/).
The systemic velocities are in agreement 
with the values used by the previous studies:
$V_{\rm sys}\simeq 3.7 $km s$^{-1}$ for VLA 1623 \citep{kamazaki03},
3.9 km s$^{-1}$ for EL 32 \citep{kamazaki03},
5 km s$^{-1}$ for EL 29 and LFAM 26 \citep{bussmann07}, 
and 3.8 km s$^{-1}$ for IRS 44 \citep{sekimoto97}.
The angle $\xi$ is the inclination angle of an outflow,
which is generally uncertain.  
The three outflows from VLA 1623, EL 29, and LFAM26 in the present paper 
are expected to be almost parallel to the plane of the sky.
Therefore, we adopt $\xi = 80^\circ$ \citep[see][]{andre90}. 
For the remaining EL 32 and IRS 44 outflows, we adopt $\xi = 57.3^\circ$, 
following \citet{bontemps96}.
The projected maximum size of each outflow lobe, $R_{\rm obs}$, was measured at 
the 3 $\sigma$ contour in its channel maps.
The maximum size of each outflow lobe was estimated by correcting for
the projection with the inclination angle $\xi$ 
as $R_{\rm out} = R_{\rm obs}/\sin \xi$.
The velocity range of each outflow was determined using the channel maps:
isolated components from high-velocity CO emission are clearly
identifiable at more than 3 $\sigma$ level within the velocity range,
without significant contamination of emission from the ambient gas.
We also note that for all the previously-identified outflows, 
the velocity ranges for the integration are more or less similar to 
those listed in Tables \ref{tab:co32} and \ref{tab:co10}.

From these quantities, we calculated the characteristic velocity
$V_{\rm out} = P_{\rm out}/M_{\rm out}$, 
dynamical time $t_{\rm dyn}=R_{\rm out}/V_{\rm out}$,
mass loss rate $\dot{M}_{\rm out}=M_{\rm out}/t_{\rm dyn}$, 
outflow luminosity $L_{\rm out}=E_{\rm out}/t_{\rm dyn}$, and 
outflow momentum flux $ F_{\rm out} = P_{\rm out}/t_{\rm dyn}$.
Tables \ref{tab:co32} and \ref{tab:co10} summarize the outflow parameters 
derived from the CO ($J=3-2$) and CO ($J=1-0$) data, respectively.
The CO lines often become optically thick even toward the outflow 
wing components. Therefore, the physical parameters such
as the outflow mass and momentum listed in Tables \ref{tab:co32} 
and \ref{tab:co10}, where both the emission lines 
are assumed to be optically thin, give the lower limits.

Most of the physical quantities estimated from CO ($J=3-2$) and 
CO ($J=1-0$) are in reasonably good agreement with each other.
The total outflow luminosity estimated from CO ($J=3-2$)
is only about 6 \% larger than that of CO ($J=1-0$).
However, for the EL 32 outflow, the mass estimated from CO ($J=1-0$) 
is larger by a factor of 4 than that of CO ($J=3-2$). 
This deviation causes large discrepancy in the other quantities 
such as the momentum, momentum flux, and luminosity.
This deviation on the estimated outflow mass is greatest 
among those of the five outflows. 
This may imply that the EL 32 outflow is more evolved 
and thus the outflow gas density may be smaller than
the critical density of the CO $(J=3-2)$ transition, 
$3\times 10^4$ cm$^{-3}$ .
If this is the case, we are probably underestimate 
the CO ($J=3-2$) outflow mass because of the LTE assumption.
This interpretation is consistent with the fact that 
the dynamical time of the EL 32 outflow is long 
and that the outflow lobes are most spread out 
with the widest opening angles.

\section{Role of Protostellar Outflows in Clustered Star Formation}
\label{sec:discussion}

\subsection{Interaction between the Outflows and Dense Gas in $\rho$ Oph}

Recent observations suggest that the dynamical interaction of
dense gas with protostellar outflows is likely to trigger 
the formation of dense cores. For example, \citet{sandell01} 
performed the 850 $\mu$m dust continuum emission observations 
toward a nearby pc-scale cluster forming clump, 
NGC 1333, and found that the dense cores are distributed 
along multiple shells and filaments that are most likely 
created by protostellar outflows.
Recently, \citet{shimajiri08} observed the Orion Molecular Cloud-2 FIR
3/4 regions using the Nobeyama Millimeter Array, and 
revealed that the $^{12}$CO emission toward the FIR 4 region 
shows an L-shaped structure in the position-velocity diagram, suggesting
that an outflow from the FIR 3 region compressed the gas in the FIR 4 region,
triggering the formation of multiple dense cores 
in the past few $\times 10^4 $ yr.
These observations suggest that dynamical compression due to 
protostellar outflows regulates the formation of dense cores 
 in pc-scale cluster forming clumps.  
Here, we discuss how the outflows in $\rho$ Oph affect
the structures of the dense gas. 

Using the H$^{13}$CO$^+$ ($J=1-0$) data obtained with the Nobeyama 
45 m telescope, \citet{maruta10} identified 68 dense cores in 
the central dense region of the $\rho$ Ophiuchi main cloud.
The H$^{13}$CO$^+$ ($J=1-0$) transition has a critical 
density of about $10^5$ cm$^{-3}$, and therefore is suitable 
as a dense gas tracer. In the following, we compare the H$^{13}$CO$^+$ data
with the outflows identified in the present paper, 
to examine whether the outflows are interacting with the dense gas 
in this region.

\begin{figure}
\epsscale{0.6}
\plotone{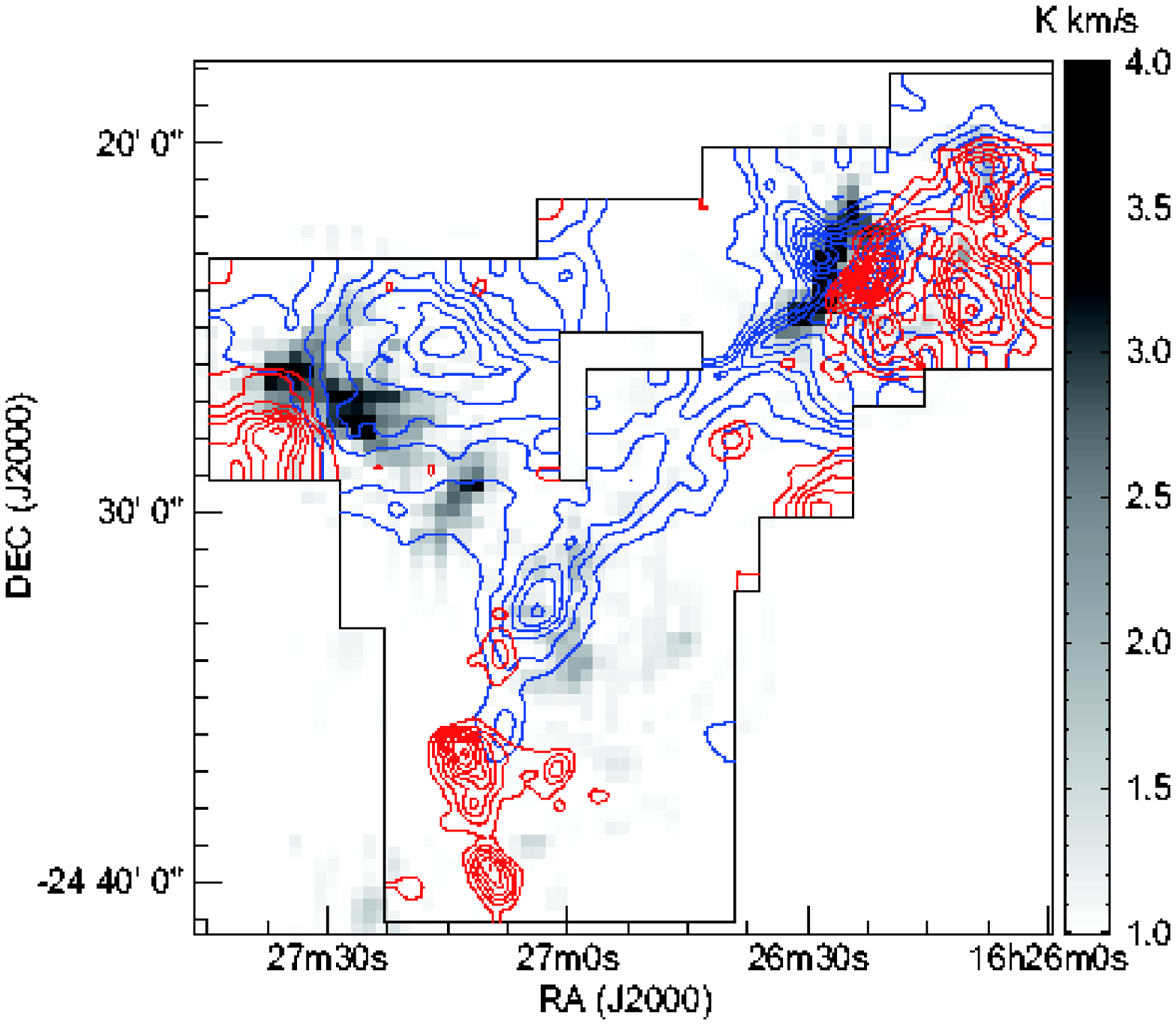}
\plotone{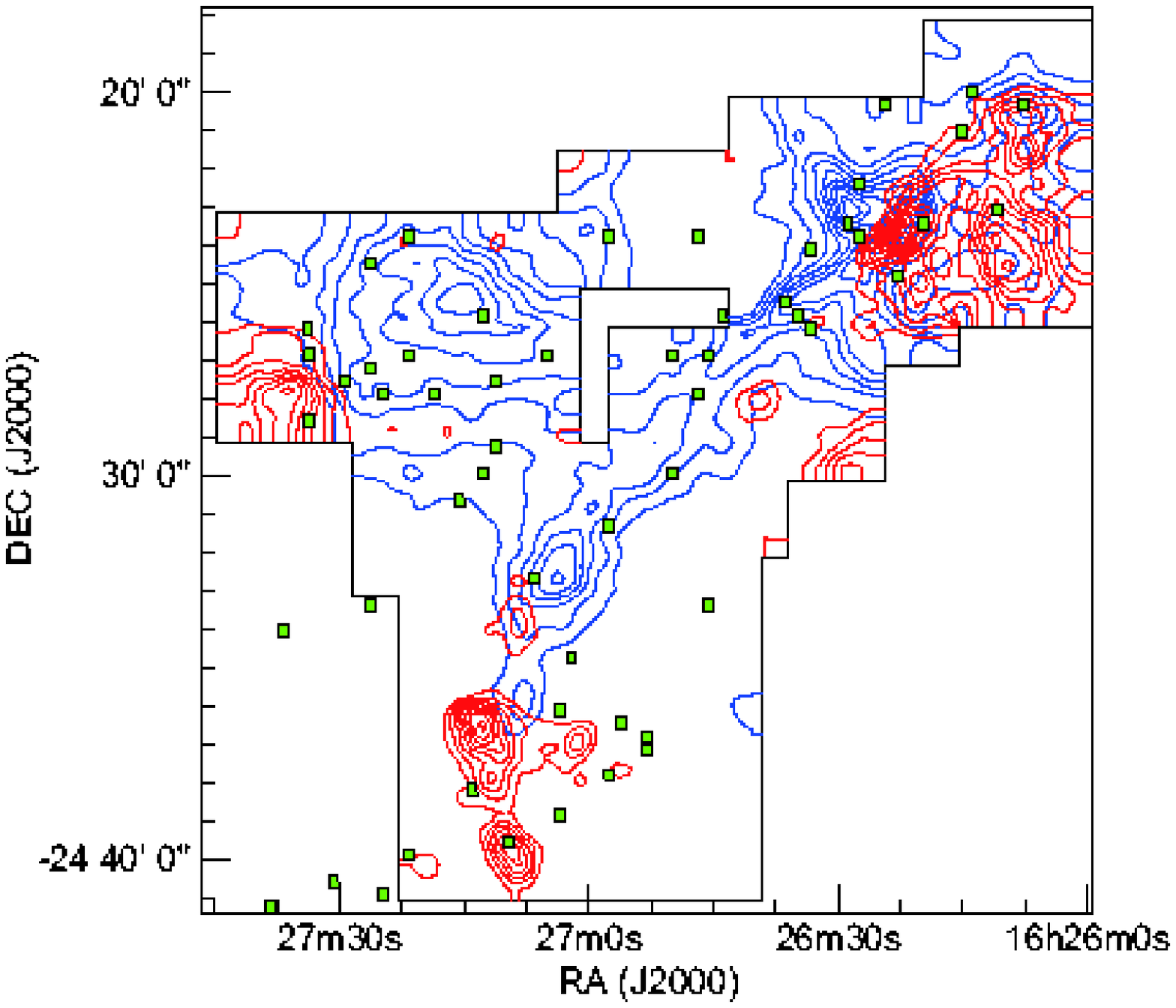}
\caption{
(a) CO ($J=3-2$) map of the $\rho$ Ophiuchi main cloud
overlaid on the H$^{13}$CO$^+$ ($J=1-0$) integrated intensity map
obtained by \citet{maruta10} in gray scale.
(b) CO ($J=3-2$) map of the $\rho$ Ophiuchi main cloud.
The positions of the H$^{13}$CO$^+$ cores 
 identified by \citet{maruta10} are overlaid by the squares.
For both the panels (a) and (b), the blue and red contours are the same 
as those of Fig. \ref{fig:map1}.
}  
\label{fig:map3}
\end{figure}

The interaction between the CO outflows and the H$^{13}$CO$^+$ dense gas 
seems distinct in the Oph B2 region. In Figure  \ref{fig:map3}a,
we present the CO ($J=3-2$) images on the H$^{13}$CO$^+$ total
integrated intensity map. We also show the positions of the
H$^{13}$CO$^+$ cores identified by \citet{maruta10} 
in Figure \ref{fig:map3}b by open squares.
About 30 out of 50 H$^{13}$CO$^+$ cores located in our 
CO ($J=3-2$) observed area are distributed around 
the outflow lobes. 
For example, in the Oph B2 region, the dense region detected by the
H$^{13}$CO$^+$ emission has an arc-like shape, 
following the south-east part of the blueshifted lobe.
The part with strong CO ($J=3-2$) emission 
in the blueshifted lobe appears to be less dense:
the peak of the CO ($J=3-2$) blueshifted emission does not 
correspond to any peak of the H$^{13}$CO$^+$ emission.

\begin{figure}
\epsscale{0.7}
\plotone{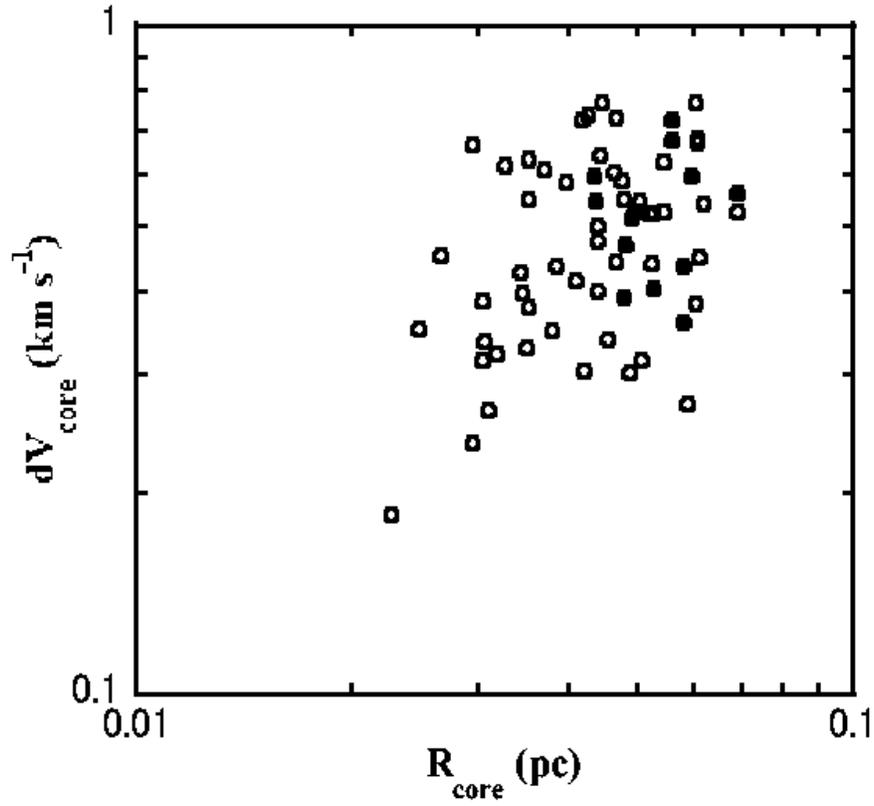}
\caption{
Line-width-radius relation of the H$^{13}$CO$^+$ cores identified by
 \citet{maruta10}. Filled circles are the cores that appear to
be associated with the EL 32 outflow lobes, while open circles
the other cores. 
}  
\label{fig:line-width-size}
\end{figure}

The EL 32 outflow possibly injected some energy into the H$^{13}$CO$^+$ 
dense cores.
In Figure \ref{fig:line-width-size}, 
we present the line-widths of the H$^{13}$CO$^+$ cores against
their radii, which is essentially the same as 
Figure 9a of \citet{maruta10}. 
The H$^{13}$CO$^+ $ cores associated with the Oph B2 region,
which are indicated by the filled circles, tend to have 
larger velocity widths.
This tendency seems to be consistent with the observations by 
\citet{friesen09}, who performed NH$_3$ observations toward the 
B1, B2, C, and F regions and found that the NH$_3$ cores in 
the B2 region have somewhat larger velocity widths. In addition, they 
found that the Oph B2 region is slightly warmer ($T\sim 15$ K)
than the other regions ($T\sim 12$ K).
\citet{friesen10} also revealed that the 
ratio of the NH$_3$ abundance to N$_2$H$^+$ abundance 
in the Oph B2 region 
appears to be smaller than that of the other regions.
These facts appear to support the idea that 
the Oph B2 region is largely affected by the outflow activity. 
However, there is a possibility that the cores in the B2 region
simply tend to follow the line width-radius relation for which
the cores with larger radii have larger line widths,
although the line-widths of the H$^{13}$CO$^+$ cores seem
to be almost independent of the core radii for $\rho$ Oph
\citep{maruta10}.
We need further observational evidence to 
quantify the interaction of the dense gas with the outflow.

In addition to the EL 32 outflow, the VLA 1623 outflow also seems
 to interact with the dense gas. 
The head of the collimated blueshifted lobe of the VLA 1623 outflow 
is located just between the Oph B1 and C regions, which might
be broken up into the two parts by the outflow.
The No. 33 H$^{13}$CO$^+$ core identified by \citet{maruta10} 
appears to collide with the VLA 1623 outflow near the edge of the Oph C region.
There are also about 10 H$^{13}$CO$^+$ cores that are distributed around 
the collimated blueshifted lobe of the VLA 1623 outflow.

The redshifted lobe of the EL 29 outflow has an S-shape axis, 
going through the less dense area between the Oph E 
and F regions. This suggests that the outflow was presumably 
bent into an S-shape as a result of the dynamical interaction with the 
dense filamentary ridge. 
The redshifted lobe has the two peaks labeled (h) and (i),
 near which two H$^{13}$CO$^+$ cores 
(No. 34 and 39 in Table 1 of \citet{maruta10}) are located.  
The difference between the radial velocity of the cores 
($\simeq 4 - 5$ km s$^{-1}$) and velocity of the redshifted lobe 
ranges from 2 and 3 km s$^{-1}$. Thus, the external pressure
exerted on the core surfaces is expected to be significant.
In fact, the cores have the virial ratios of 3 $-$ 6, suggesting 
that they are not gravitationally confined but
 compressed by the external pressure.
Furthermore, the detailed virial analysis performed by \citet{maruta10}
indicates that most of the cores in $\rho$ Oph seem not 
to be confined by their self-gravity, but to be confined 
by the ambient turbulent pressures. 
This trend is in good agreement with 
recent numerical simulations of cluster formation by
Nakamura \& Li (2010, in preparation).
These characteristics appear to be consistent with the idea 
that the outflows trigger the formation of dense cores 
in the central part of $\rho$ Oph.
However, the core distribution projected on the plane of the sky 
may be just a coincidence. In the 3D space, the cores may not 
interact with the outflow lobes.  
To directly verify the interaction with the outflows, 
we need additional observations such as using 
SiO and CH$_3$OH emission, known as shock tracers.

The importance of the external pressures
on core dynamical states has been suggested even in 
the nearby low-mass star forming regions such as the Pipe nebula
\citep{lada08}.  \citet{lada08} performed a detailed virial analysis
of the cores identified from the extinction map of this region, and 
found that all the cores have virial ratios larger than unity,
suggesting that most of the cores are pressure-confined.
However, it is unlikely that the large external pressure
is due to the protostellar outflows because star formation 
activity is quite low in the Pipe nebula
\citep{forbrich09}.
Furthermore, the typical external pressure of the Pipe nebula 
($P/k_B \sim 8\times 10^4$ K cm$^{-3}$) is nearly an 
order of magnitude smaller than that of $\rho$ Oph 
($P/k_B \sim 3\times 10^6$ K cm$^{-3}$), where
$k_B$ is the Boltzmann constant.
\citet{lada08} proposed that the source of the 
external pressure is due to the self-gravity of the Pipe nebula itself.  
In any case, the previous and our studies indicate that 
the ambient turbulent pressures of the molecular clouds 
play a significant role in core dynamics 
for both clustered and distributed star formation.

\subsection{A Dynamical Role of the Outflows in $\rho$ Oph}

\subsubsection{Turbulent Dissipation and Generation}
\label{subsub:turbulent dissipation}

On the basis of  numerical simulations, 
\citet{maclow99} derived the approximated formula 
for the energy dissipation rate of supersonic turbulence as
\begin{equation}
L_{\rm turb} \simeq f \frac{1/2 M \Delta V^2}{\lambda_d/\Delta V}
\end{equation}
where $f (\simeq 0.33)$ is the nondimensional coefficient,
$\Delta V$ is the 1D FWHM velocity width, 
and $\lambda_d$ is the driving scale of 
supersonic turbulence (see equation [7] of \citet{maclow99}).
Applying this formula,  we evaluate the rate 
of turbulent energy dissipation in the $\rho$ Ophiuchi main cloud.
\citet{loren89} estimated the total gas mass of the whole 
$\rho$ Ophiuchi main cloud (corresponding to the R21, R22, R24, R25, and R26
regions in his Table 1) to be 883 $M_\odot$ using the $^{13}$CO
($J=1-0$) emission, where the mass is recalculated by 
using our assumed distance of 125 pc to the cloud.
The 1D FWHM velocity width and the mean diameter of $\rho$ Oph (L1688)
are estimated to be $\Delta V \simeq 1.5$ km s$^{-1}$
and about 1.6 pc, respectively \citep[see][]{loren89}.
The rate of the turbulent energy dissipation is estimated to be 
\begin{equation}
L_{\rm turb} \simeq 0.03 L_\odot  \ ,
\end{equation}
where we took the cloud diameter as the driving scale of the turbulence.
We note that the driving scale of the outflow-driven turbulence 
is rather uncertain because the driving scale may vary significantly
between outflows. In fact, the size of the outflows detected 
from our observations range from 0.1 pc to 0.6 pc.
For example, \citet{maury09} adopted the cloud 
diameter as the driving scale of the outflow-driven turbulence. 
On the other hand, \citet{arce10} adopted the driving scale 
of 0.2 pc for NGC 1333, although they mentioned that this may be a lower
limit because  the most outflow lobes they identified in NGC 1333
have larger sizes.    From a theoretical consideration,
\citet{matzner07} derived the driving scale of 0.4 pc for the
outflow-driven turbulence.  
\citet{nakamura07} obtained a driving scale similar to that of 
\citet{matzner07}
on the basis of the numerical simulations of cluster formation. 
If we adopt $\lambda_d = 0.4$ pc, then the energy dissipation 
rate increases up to $L_{\rm turb} \simeq 0.12 L_\odot$.

From the analysis in the previous section,
we can evaluate the total energy injection rate 
(mechanical luminosity) due to the protostellar outflows in L1688
as
\begin{equation}
L_{\rm tot} \simeq \sum _k{E_{\rm out, {\it k}} 
\over t_{\rm dyn,{\it k}}} \simeq 0.2 L_\odot   \ .
\end{equation}
where $E_{\rm out, {\it k}}$ and $t_{\rm dyn, {\it k}}$ are
the outflow energy and dynamical time of the $k$-th outflow,
respectively, and we used the quantities obtained from 
the CO ($J=3-2$) observations.
Here, we also included the contribution from small
outflows detected in previous observations 
\citep{bontemps96,sekimoto97,kamazaki03,stanke06}.
The physical parameters of such outflows are 
listed in Table \ref{tab:others}, where 
all the quantities have been scaled such that they correspond
to an assumed distance of 125 pc.
The estimated $L_{\rm tot}$ is sensitive to the 
inclination angles assumed. If we adopt $\xi=57.3^\circ$ 
(random orientation) for all the outflows, the above value is reduced to 
$L_{\rm tot}\simeq 0.03L_\odot$.
The estimated $L_{\rm tot}$ gives a lower limit of the outflow energy 
injection rate because we assumed the optically thin for the 
CO line emission and we neglected the contribution 
from undetected faint outflows.
Therefore, we conclude that the total outflow energy injection rate is 
larger than or at least comparable to the 
turbulent dissipation rate.  

\subsubsection{Force Exerted by the Outflows}
\label{subsub:force}

To clarify how the outflows affect the dynamical state of the cloud, 
we assess the force balance in the cloud, following  \citet{maury09}.
To prevent the global gravitational contraction, the following 
pressure gradient is needed to achieve the hydrostatic equilibrium:
\begin{equation}
{dP_{\rm grav} \over dr} \simeq -G{M(r) \rho (r) \over r^2} \left(1-\alpha^{-2}\right) \ ,
\end{equation}
where $M(r)$ is the mass contained within the radius $r$ and we assume that the cloud is spherical.
The effect of magnetic field is taken into account by the factor $(1-\alpha^{-2})$
and $\alpha$ is the mass-to-magnetic flux ratio normalized to the
critical value (e.g., Nakano 1998).
Assuming the density profile of $\rho \propto r^{-2}$, the pressure 
needed to support the cloud against the gravity is estimated to be
\begin{equation}
P_{\rm grav} \simeq {GM(R)^2 \over 8\pi R^4} \left(1-\alpha^{-2}\right) \ .
\end{equation}
The force needed to balance the gravitational force is thus
evaluated to be 
\begin{equation}
F_{\rm grav} \simeq 4 \pi R^2 P_{\rm grav} (R) = {GM(R)^2 \over 2 R^2} \left(1-\alpha^{-2}\right) \ .
\end{equation}
Adopting $M(R)= 883 M_\odot$, $R=0.8$ pc \citep{loren89}, 
and $\alpha=1.4$, 
$F_{\rm grav}$ can be estimated to be 
\begin{equation}
F_{\rm grav} = 1.3 \times 10^{-3} M_\odot \ {\rm km \ s^{-1} yr^{-1}} \
 ,  
\end{equation}
where we used the same $\alpha$ as that estimated 
by \citet{sugitani10} for the Serpens main core, a nearby 
cluster forming clump similar to $\rho$ Oph, although
this value is highly uncertain.
The moderately strong magnetic field of $\alpha = 1.4$ can 
reduce the gravitational force by a factor of 2.
On the other hand, the total force exerted by the outflows in this region
is estimated as 
\begin{equation}
F_{\rm tot} \simeq \sum _k {P_{\rm out,\it k} \over t_{\rm dyn,\it k}} 
 = 0.12 \times 10^{-3} 
 M_\odot \ {\rm km \ s^{-1} \ yr^{-1}}  \ .
\end{equation} 
The force due to the outflows, $F_{\rm tot}$
is one order of magnitude smaller than that needed to 
stop the global gravitational collapse, $F_{\rm grav}$. 
Therefore, the force exerted by the outflows seems not to
be sufficient for directly supporting the whole cloud 
against its gravitational collapse.
However, we assumed that the emission lines are optically thin
in deriving the outflow parameters, and therefore 
the estimated total force of the outflows gives the lower limit.
If we adopt the optical depth of $\tau\sim 5$ for the CO ($J=3-2$) 
line \citep{kamazaki03}, $F_{\rm out} \sim 0.7\times 10^{-3}$
$M_\odot$ km s$^{-1}$ yr$^{-1}$, which 
is about a half of $F_{\rm grav}$.
If this is the case, the force exerted by the outflows may 
significantly contribute to the cloud support.
The accurate estimate of the optical depth is necessary 
to clarify the dynamical role of the outflows on the cloud support.

\subsubsection{Comparison with other nearby cluster forming clumps}

Recent observations of nearby cluster forming regions 
have revealed the detailed structure of pc-scale dense clumps 
and star formation activity including the outflows. 
Here, using the same method as in Section 
\ref{subsub:force}, we compare the dynamical 
states of two nearby cluster forming clumps, NGC 1333 and Serpens, 
with $\rho$ Oph.

Table \ref{tab:nearbysf} shows the global properties of 
the nearby cluster forming clumps.
For the Serpens clump, we adopted the results of \citet{sugitani10}
who used the outflow parameters estimated by \citet{davis99}.
For NGC 1333, we estimated the physical quantities 
using the outflow parameters obtained by \citet{sandell01} and the clump mass 
obtained by \citet{ridge03}.
The outflow parameters listed in Table \ref{tab:nearbysf}
are estimated under the assumption of optically thin. Therefore, 
they give the lower limits.
Recently, for NGC 1333, \citet{arce10} performed
a detailed analysis on the effect of outflows in the parent clump
using the $^{12}$CO ($J=1-0$) and $^{13}$CO ($J=1-0$) lines,
taking into account the effect of opacity.
The virial mass is estimated from the total mass and 
mean FWHM velocity width as
\begin{equation}
M_{\rm VIR} = 210 a^{-1} 
\left(\frac{R_{\rm cl}}{\rm pc}\right) 
\left(\frac{\Delta V}{\rm km \ s^{-1}}\right)^2  M_\odot
\end{equation}
where the dimensionless parameter $a$ is set to $5/3$,
corresponding to the density profile of $\rho \propto r^{-2}$.
The cloud masses, velocity widths, and cloud radii are 
adopted from the references indicated in the caption of Table
\ref{tab:nearbysf}.
For all the clumps, the energy injection rate of the outflows 
seems larger than or comparable to the turbulent dissipation rate.
Thus, we expect that the protostellar outflows
can power supersonic turbulence in these three
pc-scale cluster forming clumps.
Our analysis also indicates that for Serpens and NGC 1333, 
the outflows can exert the significant force against global
gravitational collapse.  However, for $\rho$ Oph, 
the total force of the observed outflows appears not to be enough 
to support the whole cloud even if we correct the effect of opacity
by a factor of $\sim 5$.  
In fact, the virial ratio is quite small,
and the clump should collapse globally.
We note that the estimated virial ratio of $\rho$ Oph is 
insensitive to the adopted line. 
For example, if we use the C$^{18}$O ($J=1-0$) emission, 
the virial ratio is estimated to be about 0.2, where
we adopted the cloud mass of 494 $M_\odot$, the velocity width of 
about 1.6 km s$^{-1}$, and the radius of about 0.3 pc
from \citet{tachihara00a}.

Why is the estimated virial ratio of $\rho$ Oph so small?
One possibility is that the current star formation got less active
than before, and therefore the cloud turbulence dissipated 
significantly.  According to \citet{evans09}, the 
current star formation rate appears to be smaller than the value
averaged over the cloud lifetime for $\rho$ Oph. 
In fact, \citet{loren89b} obtained a very shallow linewidth-radius 
relation for the $\rho$ Oph clump, based on the $^{13}$CO
($J=1-0$) observations. The shallow linewidth-radius relation 
might suggest that the large-scale turbulent motions have 
dissipated significantly because of the lack of the large-scale 
energy injection events.  
If this is the case, the clump as a whole might be in the course of the 
large-scale gravitational collapse.
Another possibility 
is that the $\rho$ Oph clump as a whole
has been compressed by a large-scale shock, 
which has enhanced the cloud gravitational energy, 
reducing the virial ratio of the clump significantly.
Several previous studies suggest that the $\rho$ Oph cloud 
has been compressed by an expanding shell created by stellar winds and
supernova explosions in the Sco OB2 association, and star
formation in this cloud has  been externally triggered
\citep{vrba77, meyers85,loren86,degeus92,motte98,tachihara02}.
The cloud morphology and kinematics 
appear to support this scenario. For example, 
the south-western side of the cloud
has a sharp edge, whereas the north-eastern side has a less dense
elongated structure. This is consistent with the scenario that 
the shell passing through the cloud from the south-western side 
to the north-eastern side would compress the cloud.
The expanding shell is also observed with the HI emission.
In addition, \citet{motte98} revealed that the dense cores as well as 
several YSOs appear to be distributed along a linear chain
that is roughly perpendicular to the direction of 
shock propagation. They proposed  that the large-scale shock
has compressed the cloud to create a dense ridge that has fragmented 
into many cores to form stars.
\citet{meyers85} suggested that the velocity difference 
of $\lesssim $ 10 km s$^{-1}$ between optical absorption 
lines of CH and CH$^+$ is caused by the shock transmitted 
into the dense clump.
This yields the cloud-crushing time of $t_{cc} \sim {\rm a \ few}\ 
10^5$ yr, which is about 5 $-$ 10 times shorter than the typical age of the 
YSOs in this clump. Here, the cloud crushing time $t_{cc}$
is the timescale for the cloud to be crushed by the shock 
transmitted into the cloud.
According to numerical simulations
of the interaction of a large-scale shock with a cloud, 
the shock motions are converted into the 
cloud turbulence sufficiently within several cloud crushing times
\citep[see e.g.,][]{nakamura06}.
The large-scale shocks are sometimes considered to be 
one of the important driving sources of the cloud turbulence
\citep[e.g.,][]{mckee07}.
For $\rho$ Oph, the observed turbulent motions seem too weak
to support the whole cloud against the global gravitational collapse, 
although the shock due to the large-scale expanding shell has passed over 
some $\sim $10 $t_{cc}$ ago.
The large-scale shock seems not to have
been sufficient to supply the supersonic cloud turbulence in this clump.

Using the $^{13}$CO ($J=1-0$) data,  
\citet{padoan09} derived the power spectrum of the turbulent field
in NGC 1333 and claimed that there is no evidence
of outflow-driven turbulence in NGC 1333. This result apparently 
contradicts our conclusion. However, the $^{13}$CO emission 
can trace only tenuous inter-core gas with densities as low as 
$10^3$ cm$^{-3}$.  Thus, their power spectrum is likely to be
contaminated by the ambient components that are not associated 
with the dense cluster forming clump whose typical density
is $10^4$ cm$^{-3}$. 
In addition, it is difficult to detect protostellar
outflows from the $^{13}$CO ($J=1-0$) line because of its weak emission 
at the high-velocity wings. The line also tends to be strongly optically thick
toward cluster forming clumps. Therefore, the $^{13}$CO emission may not 
be suitable for searching the influence by protostellar
outflows in molecular clouds.  More careful analysis will be needed.
Denser molecular gas tracers such as the C$^{18}$O ($J=1-0$) line 
are likely to be more appropriate for this kind of analysis.
In fact, \citet{brunt09} investigated the turbulent field in NGC 1333
on the basis of the PCA analysis
using the $^{12}$CO, $^{13}$CO, and C$^{18}$O emission, and 
found that the power index of the turbulent spectrum derived from 
C$^{18}$O is significantly shallower than those of $^{12}$CO
and $^{13}$CO
\footnote{Recently, \citet{carroll10} performed 3D MHD simulations
of protostellar outflow-driven turbulence and demonstrated
that the measurements of the turbulence power spectrum using 
the PCA significantly underestimate the contribution of protostellar 
outflows to the power spectrum, preventing 
the detection of the characteristic driving scale of 
the outflow-driven turbulence. 
Therefore, a role of the outflow-driven turbulence
on the larger-scale turbulence in molecular clouds should be 
clarified further in future.
}, 
implying that the outflows are likely to be 
responsible for the turbulence generation in the pc-scale 
cluster forming clump \citep[see also][]{arce10}.
This result supports our conclusion.

\section{Summary}
\label{sec:summary}

We summarize the primary results of the present paper
as follows.

1. We have performed the CO ($J=3-2$) and CO ($J=1-0$) observations
toward the nearest pc-scale cluster forming clump, 
$\rho$ Ophiuchi main cloud, and identified 5 outflows, whose 
driving sources are VLA 1623, EL 32, LFAM26, EL29, and IRS 44.
We found that the EL 32 and EL 29 outflows identified by the previous
studies are larger than previously mapped.

2. We estimated the physical quantities of the outflows.
The most luminous outflow was found to be the one from
VLA 1623. We also discovered that the EL 32 outflow 
has the largest mass and momentum among the identified outflows.
The physical quantities estimated from the two lines reasonably
agree with each other.

3. We compared our outflow images with the H$^{13}$CO$^+$ map
obtained by \citet{maruta10}, and found that 30 out of 50
H$^{13}$CO$^+$ cores appear to overlap with the outflow lobes
on the plane of the sky.  
This might imply that the outflows influence the dense cores significantly.

4. We estimated the physical parameters of the identified outflows and 
measured the total outflow energy injection rate and momentum flux. 
The total outflow energy injection rate is significantly larger 
than the dissipation rate of the supersonic turbulence in this region.
We conclude that the outflows can power the supersonic turbulence in
this region.

In contrast, the outflows do not have enough 
momentum flux to support the whole clump against the global
gravitational contraction. However, this result is sensitive to
the assumed optical depth for the CO lines.  If the CO lines are 
optically thick, then the contribution of the outflow force to the 
cloud support is expected to be significant. The accurate estimate 
of the optical depth for the CO lines toward the outflows
is needed for further clarification of the role of 
the outflows on the cloud support.

5. Using the literatures, we applied the same analysis to 
the two nearby pc-scale cluster 
forming clumps: Serpens and NGC 1333, and obtained 
the total energy injection rates and momentum fluxes.
Our analysis indicates that for both the cluster forming clumps, 
the outflows are likely to play a significant role in the global 
cloud support and turbulence generation.

6. The virial ratio of $\rho$ Oph is estimated to be only 0.2 even if 
the contribution of the moderately strong magnetic support 
against global collapse is taken into account.  
Such a small virial ratio might suggest that the current 
star formation is inactive, and thus energy injection from
the forming stars is not sufficient to support the clump.
Alternatively, the small virial ratio might be a result 
of the large-scale shock compression.
The observed weak cloud turbulence may suggest that the 
large-scale shock due to supernovae cannot supply the supersonic 
turbulence significantly to this cloud.
Instead, the large-scale shock may have triggered further star formation
by increasing the gravitational energy.

\acknowledgments 
This work is supported in part by a Grant-in-Aid for Scientific Research
of Japan (20540228, 22340040).

\appendix

\section{Mass of Molecular Outflows}

In this Appendix, we derive an expression for the outflow mass
calculated from the CO ($J=3-2$) and  CO ($J=1-0$) 
emission under the assumption of 
local thermodynamic equilibrium (LTE) with the populations of 
all levels characterized by a single excitation
temperature, $T_{\rm ex}$. 
The CO ($J=3-2$) and CO ($J=1-0$) emission associated with 
the molecular outflows is also assumed to be optically thin.
In this sense, our estimated outflow masses are the lower limit.

For molecular gas in LTE, the optical depth for the transition from
upper level, $J+1$, to lower level, J, is given by
\begin{equation}
\tau_\nu = {c^2 A_{J+1,J} \over 8\pi \nu^2} {2J+3 \over 2J+1} 
\left[1-\exp\left(-{h\nu \over k_B T_{\rm ex}}\right)\right] 
{n_J L \over d \nu}  \ , 
\label{eq:tau}
\end{equation}
where $c$ is the speed of light, $A_{J+1,J}$ is Einstein's A coefficient
for the transition from the upper level to the lower level, $\nu$ is the
frequency of the transition, $k_B$ is the Boltzmann constant, 
$n_J$ is the density of the molecule in the lower level, $J$,
and $L$ is the thickness along the line of sight
[$A_{10}=7.203\times 10^{-8}$ s$^{-1}$ and 
$A_{32}=2.497\times 10^{-6}$ s$^{-1}$ for CO ($J=1-0$) and 
CO ($J=3-2$), respectively.].

The column density of the molecule in the lower level $J$ is then estimated as
\begin{equation}
N_J = \int n_{J} L d\nu 
= \frac{8\pi \nu^3}{c^3 A_{J+1,J}}{2J+1 \over 2J+3} 
\left[1-\exp\left(-{h\nu \over k_B T_{\rm ex}}\right)\right]^{-1}
\int \tau_v dv \  ,
\label{eq:column density}
\end{equation}
where the relation $d\nu = (\nu/c)dv$  is used.

The column density of the molecule in the lower level $J$ is also 
expressed using the total column density, $N_{\rm tot}$, as
\begin{equation}
N_J = {2J+1 \over Z(T_{\rm ex})} N_{\rm tot} 
\exp\left[-\frac{hBJ(J+1)}{k_BT_{\rm ex}}\right] \ , 
\end{equation}
and 
\begin{eqnarray}
Z(T_{\rm ex}) &=& \sum _{J=0}^\infty (2J+1) 
\exp\left[-\frac{hBJ(J+1)}{k_BT_{\rm ex}}\right] \\
&\approx& {k_BT_{\rm ex} \over hB} 
\left[1 + {1 \over 3}{hB \over k_BT_{\rm ex}} + {1 \over 15}
\left({hB \over k_BT_{\rm ex}}\right)^2\right]   \ ,
\end{eqnarray}
where $Z$ is the partition function and $B$ is the rotational constant
($B=57.8975$ GHz for CO).
From eq. [\ref{eq:column density}], we thus obtain
the expression for the total column density as
\begin{equation}
N_{\rm tot}={8 \pi \nu^3 \over c^3 A_{J+1,J} (2J+3)} Z(T_{\rm ex})
\exp\left[-\frac{hBJ(J+1)}{k_BT_{\rm ex}}\right]
\left[1-\exp\left(-{h\nu \over k_B T_{\rm ex}}\right)\right]^{-1}
\int \tau_v dv \ .
\end{equation}

The optical depth $\tau_\nu$ can be obtained from the expression for 
the antenna temperature at the frequency $\nu$:
\begin{equation}
T_A^* = \eta {h\nu \over k_B}\left[f(T_{\rm ex})-f(T_{\rm bg})\right]
\left[1-\exp\left(-\tau_\nu\right)\right] \approx
\eta {h\nu \over k_B}\left[f(T_{\rm ex})-f(T_{\rm bg})\right]
\tau_\nu  \  , 
\label{eq:antenna temperature}
\end{equation}
and
\begin{equation}
f(T)\equiv {1 \over \exp (h\nu/k_BT)-1}  \  ,
\end{equation}
where $\eta$ is the beam efficiency and $T_{\rm bg}$
is the temperature of the background radiation.
The beam dilution factor is set to unity.
We adopt the cosmic microwave background radiation
with $T_{\rm bg}=2.7$ K as the background radiation. 
Then, the value of $f(T_{\rm ex})$ is much larger 
than $f(T_{\rm bg})$ when $T_{\rm ex} \sim 30$ K, 
the adopted value in the present paper.

Therefore, we neglect the second term in the right hand side of 
eq. [\ref{eq:antenna temperature}], $f(T_{\rm bg})$.
The column density at the $i$-th grid on the $j$-th channel
is then given
using the relation $\nu = 2B(J+1)$ by 
\begin{equation}
N_{\rm tot, {\it i,j}}={8 \pi k_B\nu^2 \over h c^3 A_{J+1,J} (2J+3)} 
Z(T_{\rm ex})
\exp\left[\frac{hB(J+1)(J+2)}{2k_BT_{\rm ex}}\right]
\eta ^{-1} T_{A,i,j}^* \Delta v  \ . 
\end{equation}
Thus, the outflow mass calculated from the CO ($J=3-2$) and CO ($J=1-0$) 
emission is given, respectively, by
\begin{equation}
M_{32} = \sum _j M_{32,j} \ , 
\end{equation}
and
\begin{equation}
M_{10} = \sum _j M_{10,j} \ , 
\end{equation}
where 
\begin{eqnarray}
M_{32, j} &=& \mu m_{\rm H} X_{\rm CO}^{-1} \Omega D^2
\sum _i N_{\rm tot,\it i,j} \nonumber \\
&=& 9.4\times 10^{-10} \left({X_{\rm CO} \over 10^{-4}}\right)^{-1}
\left({D \over 125 {\rm pc}}\right)^2
\left({\Delta \theta \over {\rm arcsec}}\right)^2
\left({\eta_{32} \over 0.35}\right)^{-1}
\left({\sum _i T_{A,i,j}^* \Delta v \over {\rm K \ km \ s^{-1}}}\right)
 \nonumber \\
&&\times T_{\rm ex} \exp\left[{16.6 \ {\rm K} \over T_{\rm ex}}\right]
\left[1 + {1 \over 3}\left({T_{\rm bg}\over T_{\rm ex}}\right) 
+ {1 \over 15} \left({T_{\rm bg}\over T_{\rm ex}}\right)^2\right] M_\odot
 \ ,
\end{eqnarray}
and
\begin{eqnarray}
M_{10, j} &=& 
9.2\times 10^{-9} \left({X_{\rm CO} \over 10^{-4}}\right)^{-1}
\left({D \over 125 {\rm pc}}\right)^2
\left({\Delta \theta \over {\rm arcsec}}\right)^2
\left({\eta_{10} \over 0.32}\right)^{-1}
\left({\sum _i T_{A,i,j}^* \Delta v \over {\rm K \ km \ s^{-1}}}\right) 
 \nonumber \\
&&\times T_{\rm ex} \exp\left[{5.53 \ {\rm K} \over T_{\rm ex}}\right]
\left[1 + {1 \over 3}\left({T_{\rm bg}\over T_{\rm ex}}\right) 
+ {1 \over 15} \left({T_{\rm bg}\over T_{\rm ex}}\right)^2\right] M_\odot
 \ .
\end{eqnarray}
Here, $i$ is the grid index on the $j$-th channel, 
$\eta_{32}$ and $\eta_{10}$ are the main beam efficiencies of 
the ASTE and Nobeyama 45 m telescopes, respectively, 
$m_{\rm H_2}$ is the mass of a hydrogen molecule, 
$X_{\rm CO}$ is the fractional abundance of CO relative to 
H$_2$, and $\mu$ is the mean molecular weight of the gas
and is set to 2.33.
$\Omega$ is the solid angle of the object and $D$ is the
distance to the object.

\clearpage

\begin{deluxetable}{llll}
\tabletypesize{\scriptsize}
\tablecolumns{4}
\tablecaption{Molecular Outflows Identified in the Present Paper}
\tablewidth{\columnwidth}
\tablehead{\colhead{Name\tablenotemark{a}} 
& \colhead{Region} 
&\colhead{Reference\tablenotemark{b}}  &\colhead{Characteristics} 
}
\startdata
VLA 1623 & Oph A & 1,2,3,6,7 & Class 0, highly-collimated,  
strong blueshifted lobe \\
EL32 & Oph B2 & 1,7 & 
Class I, very extended  lobes, most massive and most luminous \\
EL29 & Oph E & 1,3,4,5,8 & Class I \\
LFAM 26 & Oph E & 1,4 & Class I \\
IRS 44  & Oph F & 1,4,8 & Class I \\
\enddata
\tablenotetext{a}{Names of driving sources}
\tablenotetext{b}{(1) This work; (2) \citet{andre90};
 (3) \citet{bontemps96}; (4) \citet{bussmann07}; 
(5) \citet{ceccarelli02}; 
(6) \citet{dent95}; (7) \citet{kamazaki03};
(8) \citet{sekimoto97}}
\label{tab:outflow}
\end{deluxetable}

\begin{deluxetable}{lllllllllll}
\tabletypesize{\scriptsize}
\rotate
\tablecolumns{11}
\tablecaption{Molecular Outflow Parameters Obtained From 
CO ($J=3-2$)}
\tablewidth{1.25 \columnwidth}
\tablehead{\colhead{Name} 
 &\colhead{Velocity Range}  
&\colhead{$V_{\rm out}$}  &\colhead{$R_{\rm out}$}  &
\colhead{$M_{\rm out}$} & \colhead{$P_{\rm out}$} 
& \colhead{$t_{\rm dyn}$}  &\colhead{$\dot{M}_{\rm out}$}  &
\colhead{$F_{\rm out}$}  & \colhead{$L_{\rm out}$} & 
 \colhead{Position} \\
 \colhead{} &\colhead{(km s$^{-1}$)}  
&\colhead{(km s$^{-1}$)}  &\colhead{(pc)}  &
\colhead{($10^{-2}M_\odot$)}  & \colhead{($M_\odot$km s$^{-1}$)} 
& \colhead{($10^4$ yr)}  &\colhead{($M_\odot$yr$^{-1}$)}  &
\colhead{($M_\odot$km s$^{-1}$yr$^{-1}$)}  &
\colhead{($10^{-2}L_\odot$)} & \colhead{}
}
\startdata
Blue Lobe  & & &  & & & & & &  & \\
\hline
 VLA 1623 & $-5.0 \sim 1.0 $ & 21.7 & 0.51 & 2.2 & 0.49
 & 2.3 & $9.8\times 10^{-7}$ & $2.1\times 10^{-5}$ & 4.2 & S-E \\
 VLA 1623 & $-5.0 \sim 1.0 $ & 20.4 & 0.098 &0.55 & 0.11
 & 0.47  & $1.2\times 10^{-7}$ &$2.4\times 10^{-5}$ & 4.3 & N-W \\
 EL 32 & $-5.0\sim 0.5$ &9.73& 0.25 & 2.3 & 0.22 &
 2.6 & $8.8\times 10^{-7}$ & $8.6\times 10^{-6}$ & 0.71  & N-W \\
 LFAM 26 & $-5.0\sim 1.0$ & 29.9 &0.089 & 0.022 &0.0065
 & 0.29 & $7.5\times 10^{-8}$ & $2.2\times 10^{-6}$ & 0.73  & W \\
 EL 29 & $-5.0\sim 1.0$ & 21.6& 0.083 & 0.11 & 0.025
& 0.38 & $3.0\times 10^{-7}$ & $6.5\times 10^{-6}$ & 1.3  & N \\
\hline
Red Lobe &  & & &  & & & & &  &  \\
\hline
 VLA 1623 & $6.5\sim 13.0$ & 22.7 &0.096&0.25 & 0.056 
&0.41 & $6.0\times 10^{-7}$ & $1.4\times 10^{-5}$ & 2.9 & N-E \\
 EL 32 & $6.5\sim 10.0$ &5.7 & 0.17 &0.50 &0.028 &
2.9 & $1.7\times 10^{-7}$ & $9.7\times 10^{-7}$ & 0.049 & S-E \\
 LFAM 26 & $6.5\sim 12.0$ &19.4 &0.11 & 0.25 & 0.048 
& 0.55 & $4.5\times 10^{-7}$ & $8.6\times 10^{-6}$ & 1.5 & E \\
 LFAM 26 & $7.0\sim 12.0$ & 24.2 & 0.0089 
& 0.037 &  0.089 & 0.36 & $1.0\times 10^{-7}$ & 
$2.5\times 10^{-6}$ & 0.55 & W \\
 EL 29 & $6.5\sim 12.0$ & 21.6 & 0.13 & 0.22 & 0.047
& 0.60 & $3.7\times 10^{-7}$ & $8.0\times 10^{-6}$ & 1.5  & S \\
\enddata
\tablecomments{The CO ($J=3-2$) emission lines are assumed to be
 optically thin. Therefore, the physical quantities listed here give
the lower limits.}
\tablenotetext{a}{Relative Position of the lobe from the driving source:
N, S, E, and W mean north, south, east, and west, respectively.
For example, N-E denotes the lobe flowing from the driving source
toward north-east direction.}

\label{tab:co32}
\end{deluxetable}

\begin{deluxetable}{lllllllllll}
\tabletypesize{\scriptsize}
\rotate
\tablecolumns{11}
\tablecaption{Molecular Outflow Parameters Obtained From 
CO ($J=1-0$)}
\tablewidth{1.25 \columnwidth}
\tablehead{\colhead{Name} 
 &\colhead{Velocity Range}  
&\colhead{$V_{\rm out}$}  &\colhead{$R_{\rm out}$}  &
\colhead{$M_{\rm out}$} & \colhead{$P_{\rm out}$} 
& \colhead{$t_{\rm dyn}$}  &\colhead{$\dot{M}_{\rm out}$}  &
\colhead{$F_{\rm out}$}  & \colhead{$L_{\rm out}$} & 
\colhead{Position}  \\
 \colhead{} &\colhead{(km s$^{-1}$)}  
&\colhead{(km s$^{-1}$)}  &\colhead{(pc)}  &
\colhead{($10^{-2}M_\odot$)}  & \colhead{($M_\odot$km s$^{-1}$)} 
& \colhead{($10^{4}$yr)}  &\colhead{($M_\odot$yr$^{-1}$)}  &
\colhead{($M_\odot$km s$^{-1}$yr$^{-1}$)}  &
\colhead{($10^{-2}L_\odot$)}  & \colhead{}  
}
\startdata
Blue Lobe  & & &  & & & & & & \\
\hline
 VLA 1623 & $-3.0\sim 1.0$ & 20.8& 0.52 & 3.9 &  0.78
 & 2.5 & $1.5\times 10^{-6}$ & $3.2\times 10^{-5}$ & 5.8 & S-E \\
 VLA 1623 & $-3.0\sim 1.0$ & 18.8& 0.084 & 0.34 &  0.065 
& 0.44 & $8.0\times 10^{-7}$ & $1.5\times 10^{-5}$ & 2.3 & N-W \\
 EL 32 & $-3.0\sim 0.5$ & 9.4& 0.22 & 10.0 &  0.94 
& 2.3 & $4.4\times 10^{-6}$ & $4.1\times 10^{-5}$ & 3.3 & N-W \\
 LFAM 26 & $-3.0\sim 1.5$ & 20.7 & 0.053 & 0.023 & 0.0050
 & 0.25 &$9.5\times 10^{-8}$ & $1.9\times 10^{-6}$ & 0.33 & E \\
 EL 29 & $-3.0\sim 1.5$ & 13.5  & 0.074 & 0.25 & 0.033
 & 0.54 & $4.7\times 10^{-7}$ & $6.2\times 10^{-6}$ & 0.69  & N \\
 IRS 44 & $-3.0\sim 1.0$ & 8.7  & 0.036 & 0.15 &0.013 
& 0.40 & $3.7\times 10^{-7}$ & $3.2\times 10^{-6}$ & 0.26  & S-W \\
\hline
Red Lobe &  & & &  & & & & &  \\
\hline
 EL 32 & $6.5\sim 9.0$ & 4.6 & 0.31 & 3.2 & 0.15 
& 6.5 & $5.0\times 10^{-7}$ & $2.3\times 10^{-6}$ & 0.087  & S-E \\
 LFAM 26 & $6.5\sim 9.0$ & 15.8 & 0.10 & 0.31&  0.048
& 0.64 & $4.8\times 10^{-7}$ & $7.6\times 10^{-6}$ & 0.99  & E \\
 LFAM 26 & $6.5\sim 9.0$ & 15.4 & 0.068 & 0.045&  0.0070
 & 0.43 & $1.0\times 10^{-7}$ & $1.7\times 10^{-6}$ & 0.21  & W \\
 EL 29 & $6.5\sim 9.0$ & 17.6 & 0.096 & 0.34 & 0.062 
& 0.54 & $6.5\times 10^{-7}$ & $9.7\times 10^{-6}$ & 1.7 & S \\
 EL 29 & $6.5\sim 9.0$ & 16.9 & 0.083 & 0.21 & 0.034
 & 0.48 & $4.3\times 10^{-7}$ & $7.2\times 10^{-6}$ & 1.0 & N \\
 IRS 44 & $6.5\sim 9.0$ & 8.4 & 0.089 & 0.14 & 0.012
 & 1.0 & $1.3\times 10^{-7}$ & $1.1\times 10^{-6}$ & 0.083 & N \\
\enddata
\tablecomments{The CO ($J=1-0$) emission lines are assumed to be
 optically thin.
Therefore, the physical quantities listed here give
the lower limits.}
\label{tab:co10}
\end{deluxetable}

\begin{deluxetable}{llllll}
\tabletypesize{\scriptsize}
\tablecolumns{6}
\tablecaption{Molecular Outflow Parameters Compiled From the Literature
in the $\rho$ Ophiuchi Main Cloud}
\tablewidth{\columnwidth}
\tablehead{\colhead{Name} 
 &\colhead{$V_{\rm out}$} 
&\colhead{$M_{\rm out}$}  &\colhead{$F_{\rm out}$}  
&\colhead{$L_{\rm out}$}  &\colhead{Reference\tablenotemark{a}} \\
\colhead{} & \colhead{(km s$^{-1}$)}  &\colhead{($M_\odot$)} 
&\colhead{($M_\odot$ km s$^{-1}$ yr$^{-1}$)} &\colhead{($L_\odot$)}  & 
\colhead{}}
\startdata
WL 6   & 7.5  & $1.5\times 10^{-3}$ &$2.0\times 10^{-6}$ 
& $1.2\times 10^{-3}$  & 1 \\
WL 10 & 7.5  & $5.7\times 10^{-3}$ & $3.5\times 10^{-6}$ 
& $2.2\times 10^{-3}$  & 1 \\
AN & 3  & $8.2\times 10^{-3}$  & $2.6\times 10^{-6}$ 
&$7.0\times 10^{-3}$  & 2 \\
AS & 2  &$2.9\times 10^{-3}$  &$3.7\times 10^{-6}$ &$2.9\times 10^{-3}$ 
 & 2 \\
MMS126 & 6.5 & $1.0\times 10^{-3}$  & $0.19\times 10^{-6}$ 
&  $0.059\times 10^{-3}$  & 3 \\
WL 12  &   &      & $1.4\times 10^{-6}$ & & 4  \\
IRS 43          &  &      & $3.4 \times 10^{-6}$ & & 4  \\
IRS 48 &   &  & $6.9\times 10^{-6}$  & & 4  \\
IRS 51 &   &  & $1.3\times 10^{-6}$ & & 4  \\
\enddata
\tablenotetext{a}{Reference: 1. \citet{sekimoto97}, 2. \citet{kamazaki03},
 3. \citet{stanke06}, 4. \citet{bontemps96}}
\tablecomments{All the quantities listed in the table are corrected to 
$D=125$ pc under the assumption that observed lines are optically thin. 
The values of outflow mechanical luminosities for WL 5, and WL 10
are estimated from the expression $L_{\rm out} = F_{\rm out} V_{\rm out}/2$.}
\label{tab:others}
\end{deluxetable}

\begin{deluxetable}{llllllllll}
\tabletypesize{\scriptsize}
\tablecolumns{10}
\tablecaption{Global Properties of Nearby Cluster Forming Clumps}
\tablewidth{1.05 \columnwidth}
\tablehead{\colhead{Name} 
 &\colhead{Distance}  &\colhead{Mass} 
&\colhead{Radius}  &\colhead{Velocity Width}  &\colhead{Virial Ratio}  
&\colhead{$L_{\rm turb}$}  &\colhead{$L_{\rm tot}$}  &
\colhead{$F_{\rm grav}$}  & \colhead{$F_{\rm tot}$} \\
\colhead{} &\colhead{(pc)} & \colhead{($M_\odot$)}  &\colhead{(pc)} 
&\colhead{(km s$^{-1}$)}  &\colhead{}  
&\colhead{($L_\odot$)}  
&\colhead{} 
&\colhead{($M_\odot$ km s$^{-1}$ yr$^{-1}$)} 
&\colhead{} 
}
\startdata
Serpens $^1$ & 260 & 210 & 0.46 & 2.0 & 1.1  & 0.14 $-$ 0.32
& 1.3 & $1.9 \times 10^{-4}$
&5.2$\times 10^{-4}$  \\
$\rho$ Oph $^2$ & 125 & 883 & 0.8 & 1.5 & 0.22 & 0.06 $-$ 0.12 & 0.2
 & $ 1.3\times 10^{-3}$
 & $1.2\times 10^{-4}$ \\
NGC 1333 $^3$ & 220 & 573 &0.44  &2.8 & 0.76 & 0.41 $-$ 0.90& 0.65 &$1.8\times 10^{-3}$ & $6.0\times 10^{-4}$ \\
\enddata
\tablecomments{Reference: $^1$\citet{sugitani10}, $^2$this paper, 
$^3$\citet{knee00}.
For all the clumps, the contribution of magnetic support is taken 
into account for estimation of $F_{\rm grav}$ 
whose value is reduced by a factor of 
about 2.
See \citet{sugitani10} in detail.
The outflow parameters listed in the table are the lower limits
because they are estimated under the assumption 
that the lines are optically thin. 
}
\label{tab:nearbysf}
\end{deluxetable}

\clearpage


%
%

\end{document}